\pdfoutput=1
\documentclass[10pt,letterpaper,aps,prd,notitlepage,tightenlines,nofootinbib,longbibliography]{revtex4-1}
\usepackage[english]{babel}

\usepackage[left=1in,right=1in,top=1in,bottom=1in]{geometry}
\usepackage{microtype}
\usepackage[utf8]{inputenc}
\usepackage{amsmath}
\usepackage{amsfonts}
\usepackage{amssymb}
\usepackage{amsthm}
\usepackage{braket}
\usepackage{bbm}
\usepackage{hyperref}
\usepackage{graphicx}
\usepackage{booktabs}

\renewcommand{\vec}{\mathbf}
\newcommand{\eder}{\mathrm{d}}
\newcommand{\transpose}{\mathrm{t}}
\newcommand{\SLTR}{\mathrm{SU}(1,1)}

\newcommand{\spinnet}[1]{\;\vcenter{\hbox{\includegraphics[scale=1.1]{images/#1}}}\;}
\DeclareMathOperator{\Hom}{Hom}
\newcommand{\bigboxplus}{\operatornamewithlimits{\vcenter{\hbox{\scalebox{2}{$\boxplus$}}}}}
\newcommand{\half}[1][1]{\tfrac{#1}{2}}
\newcommand{\shalf}[1][1]{\frac{#1}{2}} 
\DeclareFontFamily{U}{mathx}{\hyphenchar\font45}
\DeclareFontShape{U}{mathx}{m}{n}{<-> mathx10}{}
\DeclareSymbolFont{mathx}{U}{mathx}{m}{n}
\DeclareMathAccent{\widebar}{0}{mathx}{"73}
\newcommand{\racah}[6]{
{\textstyle
\begin{bmatrix}
#1 & #2 & #3 \\
#4 & #5 & #6
\end{bmatrix}
}}

\newcommand{\C}{{\mathbb C}}

\newcommand{\R}{{\mathbb R}}
\newcommand{\Z}{{\mathbb Z}}

\newcommand{\cB}{{\mathcal B}}

\newcommand{\cM}{{\mathcal M}}

\newcommand{\SU}{\mathrm{SU}}

\newcommand{\SL}{\mathrm{SL}}
\newcommand{\SO}{\mathrm{SO}}

\newcommand{\be}{\begin{equation}}
\newcommand{\ee}{\end{equation}}
\newcommand{\beq}{\begin{eqnarray}}
\newcommand{\eeq}{\end{eqnarray}}
\newcommand{\bes}{\begin{eqnarray}}
\newcommand{\ees}{\end{eqnarray}}

\newcommand{\mat} [2] {\left ( \begin{array}{#1}#2\end{array} \right ) }

\newcommand{\su}{{\mathfrak{su}}}

\newcommand{\la}{\langle}
\newcommand{\ra}{\rangle}

\newcommand{\bfX}{{\bf{X}}}

\def\nn{\nonumber}

\def\te{{\widetilde{e}}}
\def\tf{{\widetilde{f}}}

\def\mone{^{{-1}}}

\def\dr{\rightarrow}

\def\ot{\otimes}
\def\one{\mathbbm{1}}

\def\demi{{\shalf}}

\def\bft{{\bf{t}}}

\def\tT{\widetilde{T}}

\def\bftt{{ \bf{\widetilde t}}}

\def\taup{{w}}
\def\bftp{{\bf{w}}}
\def\bftw{{\bf{\widetilde w}}}

\def\ov{\overline}

\theoremstyle{plain}

\begin{document}
\title{3d Lorentzian loop quantum gravity and the spinor approach}

\author{Florian Girelli}\email{fgirelli@uwaterloo.ca}
\affiliation{Department of Applied Mathematics, University of Waterloo\\Waterloo, Ontario, Canada}

\author{Giuseppe Sellaroli}\email{gsellaroli@uwaterloo.ca}
\affiliation{Department of Applied Mathematics, University of Waterloo\\Waterloo, Ontario, Canada}

\date{\today}

\begin{abstract}
We consider the generalization of the ``spinor approach'' to the Lorentzian case, in the context of 3d loop quantum gravity with cosmological constant $\Lambda=0$. The key technical tool that allows this generalization is the recoupling theory between unitary infinite-dimensional representations and non-unitary finite-dimensional ones, obtained in the process of generalizing the Wigner--Eckart theorem to $\SU(1,1)$. We use $\SU(1,1)$ tensor operators to build observables and a solvable quantum Hamiltonian constraint, analogue of the one introduced by V. Bonzom  and his collaborators in the Euclidean case (with both $\Lambda=0$ and $\Lambda\neq0$). We show that the Lorentzian Ponzano--Regge amplitude is solution of the quantum Hamiltonian constraint by recovering the Biedenharn--Elliott relation (generalized to the case where unitary and non-unitary $\SU(1,1)$ representations are coupled to each other). 
Our formalism is sufficiently general  that both the Lorentzian \textit{and} the Euclidean case can be recovered (with $\Lambda=0$).  
\end{abstract}

\maketitle


\section*{Introduction}
3d quantum gravity is a nice laboratory to explore and test some issues met in the 4d theory \cite{Carlip:1998uc}: for example, it is possible to solve the Hamiltonian constraint  and to relate loop quantum gravity (LQG) to the relevant spinfoam model. This was done in the Euclidean case, with either a vanishing or negative cosmological constant \cite{Noui:2004ja, Bonzom:2011hm, Bonzom:2011nv, Bonzom:2014bua}.

The \textit{spinorial framework to quantum gravity} is a very powerful tool to formulate and solve some questions in LQG. For example, this formalism allows to define a closed algebra of (kinematical) intertwiner observables \cite{Dupuis:2010iq}. In 4d, the simplicity constraints can be rigorously dealt with to build an Euclidean spinfoam model \cite{Dupuis:2011fz}, using this closed algebra of observables.  At the quantum level, the spinor approach makes extensive use of \textit{tensor operators} (often called also \textit{grasping operators}). The key feature of tensor operators is that their matrix elements are given in terms of Clebsch--Gordan coefficients. When dealing with $\SU(2)$, the spinor operators are conveniently realized in terms of harmonic oscillators, which can make some calculations easier. The understanding that tensor operators were in fact the general mathematical structure behind the spinor approach has been the key to unlock the formulation of LQG with a non-zero cosmological constant in the 3d case \cite{Dupuis:2013lka, Dupuis:2013haa}: they provide a quantization of  the  Hamiltonian constraint  as an operator implementing a recursion relation, whose solution is the (deformed) 6j-symbol, the amplitude of the relevant spinfoam model (Ponzano--Regge or Turaev--Viro, according to the value of the cosmological constant).

This spinorial framework was always used in the Euclidean setting, and one might wonder if this works as well in the Lorentzian scheme. In 4d, some steps have been accomplished. The Lorentzian EPRL model was constructed using spinorial tools \cite{simone}; however, the Lorentzian generalization of the Euclidean holomorphic model \cite{Dupuis:2011fz} is still not known.  The Lorentzian holomorphic model was developed at the classical level but its quantum version is  still   not yet available \cite{Dupuis:2011wy}. As a warm up, it is  interesting to test whether one can use this spinorial framework in the \textit{3d} Lorentzian case. In particular, we would like to construct the closed algebra of observables and use it to construct a Hamiltonian constraint which we could solve to recover the Lorentzian spin foam model. This is the topic of the present article.

The classical part, i.e., the spinorial description of the LQG phase space in terms of $\SU(1,1)$, is not difficult; however, similarly to the 4d case,  the construction of a quantum version is not an easy task.  Indeed, until recently, all the tensor operators for  $\SU(1,1)$ were not known. Let us describe the issue at hand. A tensor operator is a set of operators that transform as a vector in some representation of the considered Lie group or Lie algebra, here $\SU(1,1)$. Since this group is non-compact, the unitary representations are infinite-dimensional and the finite-dimensional representations (isomorphic to the $\SU(2)$ ones) are non-unitary. One could build a tensor operator transforming as a unitary representation \cite{klimyk}, but that would mean considering an infinite set of operators. Instead, we can consider a set of operators transforming as a finite-dimensional (non-unitary) representation, hence a finite set of operators. It is actually possible to realize the Lie algebra generators in terms of such a set of operators when it is acting on the discrete series  (i.e. the representations of $\SU(1,1)$ characterized by a discrete eigenvalue of the Casimir) as harmonic oscillators (see Ref. \cite{wigner_eckart} and references therein): since the representation is characterized by discrete numbers, we can use harmonic oscillators (which have a discrete spectrum) to characterize them and hence act on them.  This trick does not work if one considers a representation in the continuous series (i.e. with the a continuous Casimir eigenvalue). The recent paper \cite{wigner_eckart} solved this problem and gave the equivalent realization of the harmonic oscillators when acting on a continuous series: the key difficulty was to consider the recoupling between infinite unitary representations and finite-dimensional non-unitary representations.    This means that we are now able to quantize  the classical spinor description of the 3d Lorentzian LQG phase space  using the full machinery of  tensor operators. 

With the spinor operators at hand, we can proceed to the generalization of the spinor approach \cite{Livine:2013zha, Bonzom:2011nv} (studied in the Euclidean signature) to the Lorentzian case. The key results we focus on  are the quantization of the $\SU(1,1)$ holonomy, the (closed) algebra of spinor ``observables'' from which we can construct any observables, and finally   the construction of a solvable quantum Hamiltonian constraint. The $\SU(1,1)$ case is more subtle than the $\SU(2)$ case due to the different possible choices of unitary representations; furthermore, as mentioned earlier, we will be led to consider the recoupling between unitary and non-unitary representations. As a consequence, the  spinor ``observables'' will not  properly be  observables since they will map intertwiners defined in terms of unitary representations to intertwiners defined in terms of \textit{non-unitary} representations. Nevertheless, these observables can be used as building blocks to construct proper observables as well as a solvable quantum Hamiltonian constraint, whose solution is the Lorentzian Ponzano--Regge amplitude.

The article is organized as follows. In section I, we recall the $\SU(1,1)$ representation theory, its standard recoupling theory and the less known recouplings between unitary and (non-unitary) finite-dimensional representations. We then use these notions to construct  \emph{classical} $\SU(1,1)$ tensors (spinors, vectors, spinor representation of $\SU(1,1)$) and their quantum version, i.e., $\SU(1,1)$ tensor operators, which satisfy the Wigner--Eckart theorem \cite{wigner_eckart}.  In section II, we recall the classical picture behind 3d Lorentzian loop quantum gravity. In particular, we construct the Hamiltonian constraint, following the  method used in ref. \cite{Bonzom:2011nv}. In section III, we introduce the  notion of $\SU(1,1)$  intertwiners and Racah coefficients; since we are dealing with a non-compact gauge group, special care has to be given to them.  In Section IV, we construct the quantum Hamiltonian constraint and show how the Lorentzian Ponzano--Regge amplitude solves it.  In section V, we discuss how we can recover the results of Ref. \cite{Bonzom:2011nv}, based on $\SU(2)$, within our framework. 
%

\section{Tensors and tensor operators for \texorpdfstring{$\SU(1,1)$}{SU(1,1)}}

\subsection{\texorpdfstring{$\SU(1,1)$}{SU(1,1)} representation theory}\label{irreps}\label{recoupling_theory}
The non-compact Lie group $\SU(1,1)$ is the double cover of the  proper orthocronous Lorentz group $\mathrm{SO}_0(2,1)$, just as $\SU(2)$ is the double cover of $\SO(3)$. 
A basis for the Lie algebra $\su(1,1)$ is given by
\begin{equation}
X_0=\tfrac{1}{2}
\begin{pmatrix}
i & 0\\
0 & -i
\end{pmatrix},
\quad X_1=\tfrac{1}{2}
\begin{pmatrix}
0 && 1 \\
1 && 0
\end{pmatrix},
\quad X_2=\tfrac{1}{2}
\begin{pmatrix}
0 & -i\\
i & 0
\end{pmatrix},
\end{equation}
with commutation relations
\begin{equation}
[X_0,X_1]=-X_2,\quad [X_1,X_2]=X_0,\quad [X_2,X_0]=-X_1.
\end{equation}
$X_0$ is the generator of the subgroup $\mathrm{U}(1)\cong\mathrm{SO}(2)$, i.e. spatial rotations, while $X_1$ and $X_2$ generate boosts. As usual in physics, we will work with complexified generators
\begin{equation}
J_0:=-i X_0,\quad J_\pm:=-i X_1 \pm X_2,
\end{equation}
satisfying
\begin{equation}\label{q-su11}
[J_0,J_\pm]=\pm J_\pm,\quad [J_+,J_-]=-2J_0.
\end{equation}
The Casimir operator is given by
\begin{equation}
Q=(X_0)^2 - (X_1)^2 - (X_2)^2 \equiv -J_0(J_0+1)+J_-J_+.
\end{equation}

Complex irreducible representations of $\SU(1,1)$ fall into four different classes\footnote{We only consider here \emph{admissible} representations, i.e. those that are unitary when restricted to the maximal compact subgroup $\mathrm{U}(1)$.}. In each case, the vector space is spanned by the orthonormal vectors $\ket{j,m}$, where $j$ labels the representation and $m$ belongs to the (countable) \emph{index set} $\mathcal{M}$, which depends on the representation. The action of the generators on this vector is always the same, namely
\begin{equation}\label{transf}
\begin{cases}
J_0\ket{j,m}=m\ket{j,m}\\
J_\pm\ket{j,m}=C_\pm(j,m)\ket{j,m\pm 1}\\
Q\ket{j,m}=-j(j+1)\ket{j,m},
\end{cases}
\end{equation}
with
\begin{equation}
C_\pm(j,m)=i\sqrt{j\mp m}\sqrt{j\pm m + 1}.
\end{equation}
The possible representation classes are:
\begin{itemize}
\item {\bf Discrete series $D^\pm_j$ (positive and negative)}: infinite-dimensional representations, with
\begin{equation*}
j\in\set{-\tfrac{1}{2},0,\tfrac{1}{2},1,\dotsc},\quad \mathcal{M}^\pm=\set{\pm(j+1),\pm(j+2),\pm(j+3)\dotsc}.
\end{equation*}
They are always unitary, but only appear in the \emph{Plancherel decomposition} when $j\geq 0$.
\item {\bf Continuous series} $C^\varepsilon_j$: infinite-dimensional representations of \emph{parity} $\varepsilon\in\set{0,\frac{1}{2}}$, with
\begin{equation*}
j\in\mathbb{C},\quad \mathcal{M}=\varepsilon+\mathbb{Z}
\end{equation*}
and satisfying the constraint
\begin{equation*}
j+\varepsilon\not\in\mathbb{Z}.
\end{equation*}
The representations $C^\varepsilon_j$ and $C^\varepsilon_{-j-1}$ are isomorphic.
They are unitary only if $j\in\set{-\frac{1}{2}+is|s\neq 0}$ or, when $\varepsilon=0$, if $j\in(-1,0)$; they only appear in the Plancherel decomposition in the former case.
\item {\bf Finite-dimensional series $F_j$}: finite-dimensional representations, with
\begin{equation*}
j\in\set{0,\tfrac{1}{2},1,\dotsc},\quad\mathcal{M}=\set{-j,-j+1,\dotsc,j-1,j}
\end{equation*}
and dimension $2j+1$. The only unitary one is the \emph{trivial representation} with $j=0$, which however does \emph{not} appear in the Plancherel decomposition.
\end{itemize}

We illustrate now some results of $\SLTR$ recoupling theory. In particular we will present some non-trivial recouplings of finite and infinite-dimensional representations.

\paragraph{Coupling of finite-dimensional representations}
The finite-dimensional representations of $\SU(1,1)$ coincide with those of $\mathrm{SU}(2)$. In particular, their recoupling will have the same \emph{Cebsch--Gordan decomposition}, i.e.
\begin{equation}
F_{j}\otimes F_{j'}=\bigoplus_{J=|j-j'|}^{j+j'}F_J.
\end{equation}

\paragraph{Coupling of unitary representations}
The known recouplings for unitary representations are \cite{mukunda}
\begin{subequations}
\begin{align}
D^\pm_{j}\otimes D^\pm_{j'}&=\bigoplus_{J=j+j'+1}^\infty D^\pm_J,
\\
D^\pm_j\otimes D^\mp_{j'}&=\bigoplus_{J=J_\text{min}}^{j-j'-1}D^\pm_J \oplus \bigoplus_{J=J_\text{min}}^{j'-j-1}D^\mp_J\oplus \int_{\mathbb{R}_+}^\oplus C^\varepsilon_{-\frac{1}{2}+iS}\,\eder S,\quad J_\text{min}=\varepsilon=\varsigma(j+j'),
\\
D^\pm_j\otimes C^\varepsilon_{-\frac{1}{2}+is}&=\bigoplus_{J=J_\text{min}}^\infty D^\pm_J \oplus \int_{\mathbb{R}_+}^\oplus C^E_{-\frac{1}{2}+iS}\,\eder S,\quad J_\text{min}=E=\varsigma(j+\varepsilon),
\\
\label{eq:recoupling_CxC}
C^\varepsilon_{-\frac{1}{2}+is} \otimes C^{\varepsilon'}_{-\frac{1}{2}+is'} &= \bigoplus_{J=J_\text{min}}^\infty D^+_J \oplus \bigoplus_{J=J_\text{min}}^\infty D^-_J \oplus 2  \int_{\mathbb{R}_+}^\oplus C^E_{-\frac{1}{2}+iS}\,\eder S,\quad J_\text{min}=E=\varsigma(\varepsilon+\varepsilon'),
\end{align}
\end{subequations}
where $j,j'\geq -\tfrac{1}{2}$ and $s,s'>0$, the function $\varsigma$ is defined by
\begin{equation}
\varsigma(x)=
\begin{cases}
0\quad&\mbox{if }x\in \mathbb{Z}\\
\frac{1}{2} & \mbox{if }x\in \frac{1}{2}+\mathbb{Z},
\end{cases}
\end{equation}
all the sums are in integer steps and it is to be understood that $\bigoplus_{J=a}^b$ vanishes if $b<a$. Notice in particular that only representations in the Plancherel decomposition appear in the Clebsch--Gordan decomposition, even when we consider couplings involving discrete representations with $j=-\frac{1}{2}$. Moreover, the trivial representation $F_0$ does not appear in any of the representations. The factor $2$ in \eqref{eq:recoupling_CxC} denotes that each continuous representation appears twice in that decomposition.

\paragraph{Coupling of finite and infinite-dimensional representations} In order to make use of the full potential of tensor operators, we will need to know how the finite-dimensional representations couple with those in the discrete and continuous series; these couplings were studied in detail in \cite{wigner_eckart}. One has
\begin{equation}
F_\gamma\otimes D^\pm_j=\bigboxplus_{J=j-\gamma}^{j+\gamma}\widetilde{D}^\pm_J,
\end{equation}
with the restriction $j>\gamma-1$ and
\begin{equation}
F_\gamma \otimes C^\varepsilon_j=\bigboxplus_{J=j-\gamma}^{j+\gamma} \widetilde{C}^E_{J},\quad E=\varsigma(\gamma+\varepsilon),
\end{equation}
with the restriction that, if $j\in\mathbb{Z}/2$, $j>\gamma-1$ or $j<-\gamma$. Here $\boxplus$ denotes a direct sum only with respect to the vector space structure: the subspaces appearing in the decomposition will not be orthogonal. Moreover, we denoted the representations on the r.h.s as $\widetilde{D}$ and $\widetilde{C}$ because, with the inner product inherited from the coupling, they have a different Hilbert space structure than the usual representations: namely, the standard basis vectors $\ket{j,m}$ are orthogonal to each other but not normalized to 1. They can be brought to the standard form through a representation isomorphism which will not, however, be an isometry.
\paragraph{Clebsch--Gordan coefficients and label notation} So far  our results of recoupling theory are very heterogeneous. In order to have a uniform notation across different cases, we will introduce a new convention: the quantum number $j\in\mathbb{C}$ will become a label, i.e. we will, with abuse of notation, continue to call $j$ the pair $(j,\alpha)$, where
\begin{equation}
\alpha\in\set{D^+,D^-,C^0,C^\frac{1}{2},F}
\end{equation}
is a symbol denoting the representation class. The label $j$ now completely determines the representation, which we can denote by $\rho_j:\SLTR\rightarrow\mathrm{GL}(V_j)$, where $V_j$ is the vector space spanned by the standard basis $\ket{j,m}$. The set of possible $m$ values will be denoted by $\mathcal{M}_j$.

Consider now a generic coupling $\rho_j\otimes\rho_{j'}$. If a decomposition exists, we are going to denote by $\mathcal{D}(j,j')$ the set containing the labels of all representations appearing in it. One has
\begin{equation}
\ket{J,M}=\sum_{m,m'}A(j,m;j',m'|J,M)\ket{j,m}\otimes\ket{j',m'},\quad J\in\mathcal{D}(j,j'),\quad M\in\mathcal{M}_j,
\end{equation}
where the $A(j,m;j',m'|J,M)$'s are the \emph{Clebsch--Gordan coefficients} of the decomposition, i.e. the components of the linear map $A$ between the coupling and its decomposition. To account for the case $F_\gamma\otimes \rho_j$, in which this map is generally \emph{not} unitary, we will write
\begin{equation}
\ket{j,m}\otimes\ket{j',m'}=\int_{\mathcal{D}(j,j')} \eder\xi(J) \sum_{M\in\mathcal{M}_J}B(J,M|j,m;j',m')\ket{J,M},
\end{equation}
where the $B(J,M|j,m;j',m')$'s are the components of $A^{-1}$, which we may call \emph{inverse} Clebsch--Gordan coefficients. The integral is taken with respect to a measure $\xi$ defined as follows: let $\mathcal{D}_\alpha(j,j')\subseteq\mathcal{D}(j,j')$ denote the subset of labels with representation class $\alpha$; then
\begin{equation}
\label{eq:measure}
\xi|_{\mathcal{D}_\alpha}:=
\begin{cases}
\lambda &\quad\mbox{if }|\mathcal{D}_\alpha|=|\mathbb{R}|\\
\sum_{J\in\mathcal{D}_\alpha}\delta_J&\quad\mbox{if }|\mathcal{D}_\alpha|=|\mathbb{N}|,
\end{cases}
\end{equation}
where $\lambda$ is the Lebesgue measure and $\delta_J$ is the Dirac measure defined by
\begin{equation}
\delta_J(A)=
\begin{cases}
1\quad&\mbox{if } J\in A\\
0 &\mbox{if } J\not\in A.
\end{cases}
\end{equation}
Clebsch--Gordan coefficients possess many interesting properties. It follows from their definition that they satisfy the orthogonality relations
\begin{subequations}
\begin{equation}
\label{eq:CG_orth1}
\int\eder\xi(J)\sum_M A(j,m;j',m'|J,M)B(J,M|j,n;j',n')=\delta_{m,n}\,\delta_{m',n'}
\end{equation}
\begin{equation}
\label{eq:CG_orth2}
\sum_{m,m'} B(J,M|j,m;j',m') A(j,m;j',m'|J',M')=\delta(J,J')\,\delta_{M,M'},
\end{equation}
\end{subequations}
where 
\begin{equation}
\label{eq:delta}
\delta|_{\mathcal{D}_\alpha \times \mathcal{D}_\beta}
\begin{cases}
\mbox{is a Dirac delta}&\quad\mbox{if $\alpha=\beta$ and $|\mathcal{D}_\alpha|=|\mathbb{R}|$}\\
\mbox{is a Kronecker delta}&\quad\mbox{if $\alpha=\beta$ and $|\mathcal{D}_\alpha|=|\mathbb{N}|$}\\
\mbox{identically vanishes}&\quad\mbox{if $\alpha\neq\beta$}.
\end{cases}
\end{equation}
Moreover, they can be normalized so that
\begin{equation}
A(j,m;j',m'|J,M)\equiv B(J,M|j,m;j',m'),
\end{equation}
so that we may refer to both of them as Clebsch--Gordan coefficients.  With this normalization, they satisfy the recursion relations
\begin{equation}
\begin{split}
C_\pm(J,M)A(j,m;j',m'|J,M\pm 1)=&
C_\pm(j, m\mp 1)A(j,m \mp 1;j',m'|J,M)\\
&+C_\pm(j',m'\mp 1)A(j,m;j',m'\mp 1|J,M).
\end{split}
\end{equation}
The explicit values of some Clebsch--Gordan coefficients we will use, those of the couplings $F_\frac{1}{2}\otimes\rho_j$ with arbitrary $j$, are presented in Table \ref{CG-finite}. 

\begin{table}[ht!]
\[
\begin{array}{lccc} \toprule
& J=j-\frac{1}{2} &\quad&  J=j+\frac{1}{2}  \\ \midrule
\mu=-\frac{1}{2}\qquad &-\frac{\sqrt{j+M+\frac{1}{2}}}{\sqrt{2j+1}} && \frac{\sqrt{j-M+\frac{1}{2}}}{\sqrt{2j+1}} \\[0.8em]
\mu=+\frac{1}{2} & \frac{\sqrt{j-M+\frac{1}{2}}}{\sqrt{2j+1}} && \frac{\sqrt{j+M+\frac{1}{2}}}{\sqrt{2j+1}} \\ \bottomrule
\end{array}
\]
\caption{Values of the Clebsch--Gordan coefficient $B(J,M|\tfrac{1}{2},\mu;j,M-\mu)$ with arbitrary $j$.}
\label{CG-finite}
\end{table}

\subsection{Classical tensors} \label{cl-TO}\label{sec:tensor}
We would like to construct classical $\SU(1,1)$ tensors, where the infinitesimal action of $\SU(1,1)$ is implemented by a Poisson bracket. The first step is to realize the $\su(1,1)$ commutation relations \eqref{q-su11} at the classical level, i.e. using a Poisson bracket.
We consider therefore  the elements $x_+\in\C, \, x_0\in \R$ which satisfy  the   3d real Poisson algebra isomorphic to $\su(1,1)$
\begin{equation}
\label{cl-su11}
x_-\equiv\ov{x}_+, 
\quad \{x_+,x_-\}=2i\,x_0, \quad \{x_0,x_\pm\}=\mp i\, x_\pm.
\end{equation}
We will use these elements to generate  $\su(1,1)$ transformations, i.e. infinitesimal $\SU(1,1)$ transformations. A tensor is a set of functions that transforms as a $\SU(1,1)$ representation (or the tensor product of $\SU(1,1)$ representations). For example, we can consider the tensor $\bft^{j}_m$ which corresponds to one of the irreducible representations $\rho_j$ discussed in the previous section. As such, the tensor $\bft^{j}_m$ should transform in a similar way as \eqref{transf}, under the infinitesimal action of $\SU(1,1)$, i.e.
\begin{equation}
\label{tensor}
\{x_\pm, \bft^{j}_m\} = -i \, C_\pm(j,m)\,\bft^j_{m\pm 1} = \sqrt{j\mp m}\sqrt{j\pm m + 1}\, \bft^j_{m\pm 1}, \quad \{x_0, \bft^{j}_m\} = -i \,m\,\bft^j_{m}.
\end{equation}
Considering a unitary $\rho_j$ would mean that $\bft^{j}$ has an infinite number of components. We are going to focus instead on the finite-dimensional (non-unitary) representation $F_\gamma$, so that $\bft^\gamma$ has a finite number of components.  We will consider in particular  scalars ($\gamma=0$), spinors ($\gamma=1/2$) and vectors ($\gamma=1$). We will consider also the tensor $\bft^{\demi^*\ot\demi}$ given by   the $\SU(1,1)$ fundamental representation. To construct the latter, we need to introduce the notion of contravariant tensor, built from the dual representation $F_\gamma^*$.  Since we are dealing with finite-dimensional representations, the dual representation $F_\gamma^*$ is isomorphic to the finite-dimensional representation $F_\gamma$.  As recalled in appendix \ref{app:dual_rep}, the isomorphism is obtained as 
\begin{equation}
\bra{\gamma,\mu}\in V^*_\gamma\mapsto (-1)^{\mu}\ket{\gamma,-\mu}\in V_{\gamma}.
\end{equation}
We can therefore introduce the contravariant tensor
\be \label{contra}
\bft^{\gamma^*}_m = (-1)^{m}\bft^{\gamma}_{-m}, \quad \{x_\pm, \bft^{\gamma^*}_m\} = i \, C_\pm(\gamma,-m)\, \bft^{\gamma^*}_{m}, \quad \{x_0, \bft^{\gamma^*}_m\} = i m\, \bft^{\gamma^*}_m.
\ee
A nice property of tensors is that we can concatenate them to obtain new tensors, just like we can concatenate representations, using  Clebsch--Gordan coefficients; in fact,
\begin{equation}\label{conca}
\bft^\gamma_\mu=\sum_{\mu_1,\mu_2}A(\gamma_1,\mu_1;\gamma_2,\mu_2|\gamma,\mu)\,\bft^{\gamma_1}_{\mu_1}\bft^{\gamma_2}_{\mu_2},\quad \gamma\in\mathcal{D}(\gamma_1,\gamma_2),\quad \mu\in\mathcal{M}_\gamma
\end{equation}
is the $\mu$ component of a tensor  of rank $\gamma$, as can be checked using the recursion relations for the Clebsch--Gordan coefficients.


 We are interested in finding the ``spinor'' variables, that is a set of variables $\bft^\demi_{\pm\demi}\equiv\bft_{\pm}\in \C$ such that  
\be\label{cl-spinor}
\{x_\pm,\bft_{\pm}\}= 0, \quad \{x_\pm,\bft_{\mp}\}=  \bft_{\pm}, \quad \{x_0,\bft_{\pm}\}= \mp \half[i]\, \bft_{\pm}.
\ee
In the following, it will be convenient to consider two spinors $\bftt=\mat{c}{\bftt_-\\\bftt_+} , \, \bft=\mat{c}{\bft_-\\\bft_+} \in\C^2$.
The Poisson brackets in \eqref{cl-su11} and \eqref{cl-spinor} are realized if we set 
\begin{equation}
 x_\pm = \pm i\, \bft_\pm \bftt_\pm , \quad x_0= -\half \left( \bft_-\bftt_+ + \bft_+\bftt_-\right),
\end{equation}
with
\begin{equation}
\label{sol-t}
\{\bftt_+,\bft_-\} = \{\bft_+,\bftt_-\}=-i, \quad \{\bftt_+,\bftt_-\} = \{\bft_+,\bft_-\}= \{\bftt_+,\bft_+\}= \{\bftt_-,\bft_-\}=0.
\end{equation}
As such, we are considering a (complex) symplectic form on $\C^4\ni (\bft,\bftt)$, which is \textit{not} the canonical one.  Note that we have to implement  the reality constraints $x_-=\ov{x}_+$, and $x_0=\ov{x}_0$, so that we need to reduce our parametrization space $\C^4$ to a smaller space.    There are two natural  choices to implement the reality constraints,
\be\label{constraints}
\bft_-=\ov\bftt_+, \quad \bftt_-=\ov\bft_+ \quad \mbox{ \emph{or} }\quad  \bft_-=-\ov\bftt_+, \quad \bftt_-=-\ov\bft_+ ,
\ee
which reduce $\C^4$ to $\C^2$ equipped with the canonical symplectic form.

We can concatenate the  spinors $\bftt$ and $\bft$ to form a scalar in the following way. We use the Clebsch--Gordan coefficient $A(\half,\mu_1;\half,\mu_2|0,0)=  (-1)^{\demi-\mu} \delta_{\mu_1,-\mu_2}$ to define a bilinear form (notice that it is not diagonal, so the order we set the vectors is important) 
\begin{subequations}
\begin{align}
\cB(\bft,\bftt) &=-\sqrt{2}\,\sum_{\mu_1,\mu_2}A(\half,\mu_1;\half,\mu_2|0,0)\,\bft_{\mu_1}\bftt_{\mu_2}= (-1)^{\demi+\mu} \bft_{\mu}\bftt_{-\mu} = -\bft _+\bftt_- +  \bft _-\bftt_+ \equiv \la \tau | \tau \ra,\\
\cB(\bftt,\bft)&=-\sqrt{2}\,\sum_{\mu_1,\mu_2}A(\half,\mu_1;\half,\mu_2|0,0)\,\bftt_{\mu_1}\bft_{\mu_2}= (-1)^{\demi+\mu} \bftt_{\mu}\bft_{-\mu}= -\bftt _+\bft_- +  \bftt _-\bft_+ \equiv [ \tau | \tau]=- \la \tau | \tau \ra,\\
\cB(\bftt,\bftt) &=-\sqrt{2}\,\sum_{\mu_1,\mu_2}A(\half,\mu_1;\half,\mu_2|0,0)\,\bftt_{\mu_1}\bftt_{\mu_2}= (-1)^{\demi+\mu} \bftt_{\mu}\bftt_{-\mu} \equiv [\tau|\tau\ra=0, \\  
\cB(\bft,\bft)&=-\sqrt{2}\,\sum_{\mu_1,\mu_2}A(\half,\mu_1;\half,\mu_2|0,0)\,\bft_{\mu_1}\bft_{\mu_2}= (-1)^{\demi+\mu} \bft_{\mu}\bft_{-\mu}\equiv \la \tau|\tau]=0 .
\end{align}
\end{subequations}
We have introduced the (square) bra-ket notation as it is convenient to keep track of the nature of the spinor. The kets $| \tau\ra$ and $| \tau]$ are the initial spinors, respectively $\bftt$ and $\bft$, whereas $[\tau|$ and $\la \tau |$  are   contravariant spinors, as defined in \eqref{contra}; explicitly
\begin{equation}
| \tau\ra \equiv \bftt,\quad | \tau]\equiv \bft 
\end{equation}
and
\begin{equation}
\la \tau |\equiv((-1)^{\demi-\mu} \bft_{-\mu})_\mu= (-\bft_+,\bft_-), \quad [\tau|\equiv ((-1)^{\demi-\mu} \bftt_{-\mu})_\mu= (-\bftt_+,\bftt_-).
\end{equation}
Out of the spinor variables  $\bft$ and $\bftt$, using \eqref{conca}, we can also define a vector
\begin{equation}
\label{flux}
X_\mu= \sum_{\mu_1,\mu_2}\, A(\half,\mu_1;\half,\mu_2|1,\mu) \bft_{\mu_1}\bftt_{\mu_2} \quad\mbox{ with }\quad X_{\pm1}= \bft_\pm\bftt_\pm, \quad X_0=\frac{1}{\sqrt{2}} (\bft_-\bftt_++\bft_+\bftt_-);
\end{equation}
we can explicitly check that
\be\label{cl-vector}
\{x_\pm,X_{\pm}\}= 0, \quad \{x_\mp,X_{\pm}\}= \sqrt{2}\, X_{0} , \quad \{x_\pm,X_{0}\}=  \sqrt{2} X_{\pm}, \quad \{x_0,X_{\pm}\}= \mp i \sqrt{2}X_{\pm}, \quad \{x_0,X_{0}\}=0.
\ee
We note that the components $\bfX_i$  are related to the $\su(1,1)$ generators $x_i$ by
\be\label{sig}
X_{\pm1}= \mp i\,  x_\pm, \quad X_0= -\sqrt{2} \, x_0, \quad 
X_i= \la \tau|  \sigma_i | \tau\ra,
\ee
with
\begin{equation}
\sigma_+=\mat{cc}{0&1\\0&0},\quad  \sigma_-=\mat{cc}{0&0\\1&0},\ \sigma_0=\mat{cc}{-1&0\\0&1} \,.  
\end{equation}
By construction, having in mind the constraints \eqref{constraints}, we have that $X_+=\ov X_-$, $X_0=\ov X_0$.
The classical analogue $q$ of the $\su(1,1)$ Casimir  is proportional to the norm of $\bfX$, as
\be
|\bfX|^2 = \sqrt{3}\sum_{m_1,m_2}\, A(1,m_1;1,m_2|0,0) X_{m_1}X_{m_2}= 2X_+X_- - X_0^2 =2(x_+x_--x_0^2) = 2q.
\ee

A  $\SU(1,1)$ group element represented in   the fundamental representation is
\be
 g=\mat{cc}{\alpha&\beta\\ \overline{\beta}&\overline{\alpha}}\in \SU(1,1), \quad |\alpha|^2-|\beta|^2=1. 
\ee
It is a matrix, hence it can be seen as the tensor product of spinor and a contravariant spinor $g\sim \bft^\demi\ot \bft^{\demi^*}$. Taking into account that we must have  $\det g =1$, we consider the spinorial representation for $g$ as
\be\label{holo}
g= \frac{-i}{\sqrt{\la \tau|\tau\ra}\sqrt{ \la \taup|\taup\ra}}\left(   |\taup\ra [ \tau |- |\taup]\la \tau|  \right) = \frac{-i}{\sqrt{\la \tau|\tau\ra}\sqrt{ \la \taup|\taup\ra}}
\mat{cc}{
-\bftw_-\bftt_++\bftp_-\bft_+& -\bftp_-\bft_-+\bftw_-\bftt_-\\ 
-\bftw_+\bftt_++\bftp_+\bft_+&-\bftp_+\bft_-+\bftw_+\bftt_-}, 
\ee
with $\la \tau|\tau\ra =\la \taup|\taup\ra$. We can check that $\det g= 1$ and that $\{g_{ij}, g_{kl}\}=0$.  As an element of $\SU(1,1)$, we also require the constraints $g_{11}= \ov g_{22}$ and $g_{12}= \ov g_{21}$; if one uses the  constraints \eqref{constraints}, they are automatically satisfied. 
The inverse holonomy is given by 
\be
g\mone =  \frac{i}{\sqrt{\la \tau|\tau\ra}\sqrt{ \la \taup|\taup\ra}}\left(  |\tau\ra [ \taup| - |\tau]\la \taup| \right). 
\ee
Using the matching constraint $\la w|w\ra = \la \tau|\tau\ra$, we have 
\begin{equation}
\label{eq:g_action}
\begin{cases}
g | \tau \ra = i|w], & g | \tau ]=i|w\ra,\\
\la w|g=-i[\tau|, & [w|g=-i\la\tau|
\end{cases}
\quad \mbox{and} \quad
\begin{cases}
g\mone | w \ra = -i|\tau], & g\mone | w ]=-i|\tau\ra,\\
\la \tau|g\mone=i[w|, & [\tau|g\mone=i\la w|.
\end{cases}
\end{equation}

Using their expression in terms of the spinors, we can also calculate the Poisson bracket between the $x_i$ and the $\SU(1,1)$ matrix elements, to recover the full phase space structure of $T^*\SU(1,1)$. Hence, as expected, the spinors provide a nice parametrization of $T^*\SU(1,1)$, with
\begin{equation}
\{x_+,x_-\}=2ix_0, \quad \{x_0,x_\pm\}=\mp i x_\pm,\quad \{g_{ij}, g_{kl}\}=0, \quad \{x_\alpha,g_{ij}\}=- (g\sigma_\alpha)_{ij},
\end{equation}
where the $\sigma_\alpha$, $\alpha=\pm,0$ are given in \eqref{sig}.

%

\subsection{ Tensor operators}
\subsubsection{Definition and Wigner--Eckart theorem}\label{q-tensor}
At the representation theory level, we are interested in a set of \emph{operators} that transform as vectors in a finite-dimensional representation\footnote{The representation need not be finite-dimensional \cite{klimyk}, but we will only consider this case here.} under the action of the group. Explicitly, let
\begin{equation}
\rho_{j}:\SLTR\rightarrow \mathrm{GL}(V_{j})
\end{equation}
denote any irreducible $\SLTR$ representation, where $j$ is to be thought as a label including, in addition to its numerical value, the class of the representation (e.g. continuous, discrete positive). Given two such representations $\rho_{j_1}$ and $\rho_{j_2}$ one can associate to them a new representation
\begin{equation}
R:\SLTR\rightarrow\mathrm{GL}(\mathrm{Lin}(V_{j_1},V_{j_2}))
\end{equation}
defined by
\begin{equation}
R(g)A=\rho_{j_2}(g)A\,\rho_{j_1}(g)^{-1},\quad \forall A\in\mathrm{Lin}(V_{j_1},V_{j_2}).
\end{equation}
An (irreducible) tensor operator of rank $\gamma\in\mathbb{N}_0/2$ is an intertwiner between $R$ and the finite-dimensional representation $F_\gamma$, i.e. a linear map
\begin{equation}
T^\gamma:V_\gamma\rightarrow \mathrm{Lin}(V_{j_1},V_{j_2})
\end{equation}
such that
\begin{equation}\label{eq:tens_op_def1}
R(g) \circ T^\gamma = T^\gamma\circ F_\gamma(g),\quad \forall g\in\SLTR.
\end{equation}
As usual with linear maps, the components of a tensor operator (in a given basis) are defined by evaluating it on basis vectors: with our standard basis they are the linear maps $T^\gamma_\mu:V_{j_1}\rightarrow V_{j_2}$ defined by
\begin{equation}
T^\gamma_\mu:= T^\gamma\left( \ket{\gamma,\mu}\right),\quad\mu\in\mathcal{M}_\gamma=\set{-\gamma,\dotsc,\gamma}.
\end{equation}
Equation \eqref{eq:tens_op_def1} can be rewritten in terms of these components by evaluating both sides on each $\ket{\gamma,\mu}$, to get
\begin{equation}
\rho_{j_2}(g)\,T^\gamma_\mu\, \rho_{j_1}(g)^{-1} = \sum_{\nu\in\mathcal{M}_\gamma}\braket{\gamma,\nu|F_\gamma(g)|\gamma,\nu}T^\gamma_\nu,\quad \forall g\in\SLTR.
\end{equation}
Differentiating at the identity we obtain, at the Lie algebra level, that
\begin{equation}
\rho_{j_2}(X)\,T^\gamma_\mu - T^\gamma_\mu\,\rho_{j_1}(X)= \sum_{\nu\in\mathcal{M}_\gamma}\braket{\gamma,\nu|F_\gamma(X)|\gamma,\nu}T^\gamma_\nu,\quad \forall X\in\su(1,1),
\end{equation}
which in terms of the generators can be written in the compact form
\begin{equation}\label{eq:tens_op_algebra_cond}
[J_0,T^\gamma_\mu]=\mu T^\gamma_\mu,\quad [J_\pm,T^\gamma_\mu]=C_\pm(\gamma,\mu)\,T^\gamma_{\mu\pm 1};
\end{equation}
this can be seen as the quantum version  of \eqref{tensor}. One can show that the definitions at the group and algebra level are equivalent \cite{barut}.

Tensor operators possess two extremely useful properties. First of all, they can be combined to build other operators of greater or smaller rank: explicitly, the operator defined by
\begin{equation}
\sum_{\mu_1,\mu_2}A(\gamma_1,\mu_1;\gamma_2,\mu_2|\gamma,\mu)\,T^{\gamma_1}_{\mu_1}T^{\gamma_2}_{\mu_2},\quad \gamma\in\mathcal{D}(\gamma_1,\gamma_2),\quad \mu\in\mathcal{M}_\gamma
\end{equation}
is the $\mu$ component of a tensor operator of rank $\gamma$, as can be checked by making use of the Clebsch--Gordan recursion relations.

A second, more profound feature of tensor operators lies in the \textit{Wigner--Eckart theorem}, which roughly states that the matrix elements of a tensor operator between two irreducible representations are proportional to a Clebsch--Gordan coefficient, the proportionality factor being the same for every matrix element. Explicitly, let $T^\gamma$ be a tensor operator of rank $\gamma$ between two representations $\rho_j$ and $\rho_{j'}$. When the decomposition of $F_\gamma\otimes \rho_j$ exists, the matrix elements of the components of $T^\gamma$ are of the form
\begin{equation}
\braket{j',m'|T^\gamma_\mu|j,m}=\braket{j'\Vert T^\gamma \Vert j} B(j',m'|\gamma,\mu;j,m),
\end{equation}
where the \emph{reduced matrix element} $\braket{j'\Vert T^\gamma \Vert j}\in\mathbb{C}$ does not depend on $\mu$, $m$ or $m'$.

This theorem, originally formulated for $\mathrm{SU}(2)$, is known to be valid more generally for compact groups, and it was proved for the particular non-compact case of $\SLTR$ in \cite{wigner_eckart}. An overview of the proof of the theorem is presented in Appendix \ref{proof WE}.

\subsubsection{Spinor and vector operators for \texorpdfstring{$\SU(1,1)$}{SU(1,1)}}\label{sec:q-spinor}
We want to solve \eqref{eq:tens_op_algebra_cond} in the case $\gamma=1/2, \, 1$. We shall discuss the case $\gamma=0$ in section \ref{lqg-obs}. Due to the Wigner--Eckart theorem, we know the matrix elements of $T^\gamma$ modulo the reduced matrix element. Furthermore, just like in the classical case, we want to reconstruct the vector operator, whose components are proportional to the $\su(1,1)$,  from the spinor operators. This requirement essentially fixes the    reduced matrix element for both $\gamma=1/2$ and $\gamma=1$. 

For the spinor case, we want to solve 
\begin{equation}\label{q-spinor}
[J_0,T_\pm]=\pm \half  T_\pm,\quad [J_\pm,T_\mp]=C_\pm(\half,\mp\half)\,T_\pm, \quad [J_\pm,T_\pm]=0,
\end{equation}
which is the quantum version of \eqref{cl-spinor}. We consider two solutions $T$ and $\widetilde T$, which are characterized by the chosen reduced matrix elements 
\begin{equation}\label{ansatz2}
\braket{j'\Vert T \Vert j}=\sqrt{2j+1}\,\delta_{j',j+\demi},
\quad
\braket{j'\Vert \widetilde{T} \Vert j}=\sqrt{2j+1}\,\delta_{j',j-\demi}.
\end{equation}
As mentioned above, these normalizations are chosen so that we can recover the $\su(1,1)$ generators out of the spinor operators.  With this normalization, the action of $T$ and $\widetilde T$ on a vector $\ket{j,m}$ belonging to a (unitary) irreducible representation is 
\bes \label{spinor comp}
&& T_-\ket{j,m}=\sqrt{j-m+1}\ket{j+\tfrac{1}{2},m-\tfrac{1}{2}},\quad 
T_+\ket{j,m}=\sqrt{j+m+1}\ket{j+\tfrac{1}{2},m+\tfrac{1}{2}},\nn\\
&&
\widetilde{T}_-\ket{j,m}=-\sqrt{j+m}\ket{j-\tfrac{1}{2},m-\tfrac{1}{2}},\quad
\widetilde{T}_+\ket{j,m}=\sqrt{j-m}\ket{j-\tfrac{1}{2},m+\tfrac{1}{2}}.
\ees
It follows therefore that
\begin{equation}\label{eq:heisenberg_cr}
[T_+,\widetilde{T}_-]=[\widetilde{T}_+,T_-]=\mathbbm{1},
\end{equation}
with all other commutators vanishing. We note that the tensor operators $T$ and $\widetilde T$ might send a unitary representation to a non-unitary one. Indeed, if $j$ is a continuous representation, then $j=-\demi +is$ and $j\pm \demi$ is \emph{not} a unitary representation anymore. As we shall see later on, this means that we will consider polynomials of combinations of the type $T \widetilde T$ or $ \widetilde T T$ in order to always retrieve a unitary representation in any case. The vector operator given in terms of the generators is such an example.

Indeed, instead of $\gamma=1/2$, we can focus on the case ${\gamma=1}$ and solve \eqref{eq:tens_op_algebra_cond} in this case. It is not difficult to see that the $\su(1,1)$ generators, properly rescaled  
\begin{equation}
\label{eq:vector operator}
V_0=-\sqrt{2}J_0,\quad V_{\pm 1}=\mp i J_\pm
\end{equation}
satisfy \eqref{eq:tens_op_algebra_cond} with $\gamma=1$. $V_\mu$ is then a \textit{vector} operator. Instead of proceeding in this way, we can also recall that a vector can be built out of spinors. We use the Clebsch--Gordan coefficients to concatenate the spinor operators  to form a vector operator. 
\begin{equation}\label{eq:ansatz1}
V_\mu=\sum_{\mu_1=-\frac{1}{2}}^\frac{1}{2} \sum_{\mu_2=-\frac{1}{2}}^\frac{1}{2} \braket{\tfrac{1}{2},\mu_1;\tfrac{1}{2},\mu_2|1,\mu}
\, T_{\mu_1}\widetilde{T}_{\mu_2}.
\end{equation}
The choice of reduced matrix element for the spinor in \eqref{ansatz2} is essential to recover the generators $J_i$. We have explicitly,
\begin{equation}\label{SJ general}
J_\pm=\pm i T_\pm \widetilde{T}_\pm,\quad J_0=-\tfrac{1}{2} \left( T_- \widetilde{T}_+ + T_+ \widetilde{T}_- \right).
\end{equation}

\subsubsection{Quantization of classical tensors}\label{sec:holo}
In section \ref{cl-TO}, we have defined the notion of classical tensors. We have constructed the classical spinors  $\bft, \bftt$ and  the vector  $\bfX$. In the previous section, we have constructed the spinor and vector operators, and it is natural to ask whether they can be seen as the quantization of their classical counter-parts. It is enough to focus on the spinors, since they are the building blocks from which any other tensor is built.

At the classical level, we started with a pair of spinors $(\bft, \bftt)\in\C^4$. We built the generators $x_i$ and the holonomy $g$ out of these variables. Due to the reality constraints that both $x_i$ and $g$ have to satisfy, we had constraints on $(\bft, \bftt)$: we could therefore reduce the number of independent variables to $\C^2$. We found two natural constraints in \eqref{constraints}. 

Let us now discuss the quantization procedure. 
A priori, we have a choice: we can first implement the reality constraint, then quantize, or alternatively first quantize, then implement a quantum version of the reality constraints.  Since we have different types of operators (e.g. different adjointness properties, possibly sending unitary representations to non-unitary representations) according to the space on which they act on, positive or negative discrete series and continues series, we need to be cautious and make a study case by case. 
We first consider the case where we quantize the tensors and then find a quantum version of the necessary reality conditions. 

The (non-zero) Poisson brackets on $\C^4$ are straightforward to quantize as
\begin{equation}
\bftt\dr \widetilde T, \, \bft\dr T,\qquad  \{\bftt_+,\bft_-\}= -i  \dr [\widetilde{T}_+,T_-]=\mathbbm{1}, \qquad 
 \{\bft_+,\bftt_-\}=-i \dr [T_+,\widetilde{T}_-]=\mathbbm{1}.
\end{equation}
We consider now  the natural reality constraints  \eqref{constraints} we found. They are easily quantized as 
\begin{equation}\label{natural q constraint}
\bft_\pm=
\begin{cases}
-\ov \bftt_\mp\\
\ov \bftt_\mp 
\end{cases} \dr \quad  T_\pm=
 \begin{cases}
-\widetilde{T}^\dagger_\mp \\
\widetilde{T}^\dagger_\mp.
\end{cases}
\end{equation}
A quick look at the spinor operators action shows that these constraints are actually realized   when acting on the discrete series; we have indeed
\begin{equation}
T_\pm= \begin{cases}
-\widetilde{T}^\dagger_\mp &\quad \mbox{when acting on }D^+_j\\
\widetilde{T}^\dagger_\mp &\quad \mbox{when acting on  }D^-_j.
\end{cases}
\end{equation}
With these reality conditions, we see that we can use a pair of quantum  harmonic oscillators $(a,a^\dagger)$, $(b, b^\dagger)$ satisfying
\begin{equation}
[a,a^\dagger]=[b,b^\dagger]=\mathbbm{1},
\end{equation}
with all other commutators vanishing, to represent our spinor operators; explicitly
\begin{equation}
\begin{cases}
T=  \mat{c}{ib \\-a^\dagger} ,   \quad  \widetilde{T} =\mat{c}{a\\ib^\dagger}\quad &\mbox{for }D^+_j\\[2em]
T=  \mat{c}{-ib^\dagger \\ a} , \quad \widetilde{T} =\mat{c}{a^\dagger \\ib}  &\mbox{for }D^-_j.
\end{cases}
\end{equation}
With this parametrization, we recover the usual Schwinger--Jordan representation of the $\su(1,1)$ generators
\begin{equation}
J_+=a^\dagger b^\dagger,\quad J_-=a b,\quad J_0=
\begin{cases}
\tfrac{1}{2}\left(a^\dagger a + b^\dagger b +1\right) &\quad \mbox{for }D^+_j\\
-\tfrac{1}{2}\left(a^\dagger a + b^\dagger b +1\right) &\quad \mbox{for }D^-_j.
\end{cases}
\end{equation}

When acting on the continuous representations, things are more complicated. Indeed, in this case, the spinor operators are not related by the adjoint, but only the transpose, i.e.
\begin{equation}\label{transposed}
\braket{j+\tfrac{1}{2},m\pm \tfrac{1}{2}|T_\pm|j,m}=\mp\braket{j,m|\widetilde{T}_\mp|j+\tfrac{1}{2},m\pm \tfrac{1}{2}},
\end{equation}
since $j$ is now complex. 
One might then wonder whether, when acting on the continuous representation, one can still recover 
\be
J_+=J_-^\dagger;
\ee 
this will be true since the combinations $ i T_+ \widetilde{T}_+$ and $- i T_- \widetilde{T}_-$ have real coefficients when acting on the continuous series, as
\begin{equation}
J_\pm\ket{j,m} =\pm i T_\pm \widetilde{T}_\pm\ket{j,m} = i\sqrt{j\mp m}\sqrt{j\pm m+1}\ket{j,m\pm 1}= C_\pm(j,m)\ket{j,m\pm 1}.
\end{equation}
The coefficients $C_\pm(j,m)$ are real\footnote{Explicitly  $C_{\pm}(-\shalf+is,m)^2= s^2+1/4 +m(m\mp 1)>0, \, \forall m\in \cM=\varepsilon +\Z$.} and hence it is easy to see that we still have that $J_+=J_-^\dagger$ when expressing the generators in terms of the spinor operators.

The other possible scheme of quantization, ``reduce the reality conditions, then quantize'' would also work easily for the discrete series, when using the reality constraints \eqref{constraints}. However, for continuous representations, it is not clear how it would work, since the standard way to quantize two variables that are complex conjugate to each other is to say that their quantum versions are related through the adjoint. 
%
%

\begin{table}[h]
\begin{tabular}{ccccccc}
\toprule
            &$\qquad$ & discrete positive &$\quad$& discrete negative &$\qquad$& continuous\\ \midrule
$T_\pm =$ && $-\widetilde T_{\mp}^\dagger$ &&$\widetilde T_{\mp}^\dagger  $&& $\mp \widetilde T_{\mp}^t$ \\\bottomrule
\end{tabular}
\caption{The spinor operators $T$, $\widetilde T$ are not independent of each other. Their relations depend on the Hilbert space they act on. We use the basis $\ket{j,m}$ to define the transpose and the adjoint. }
\end{table}

To summarize, we recovered the well-known Schwinger--Jordan trick for $\su(1,1)$, previously only known for the discrete series. We have  an extension to the continuous series \cite{wigner_eckart}, but of course not in term of the usual harmonic oscillators.

The quantization of the holonomy given in \eqref{holo} has some ordering ambiguity due to the quantization of the inverse of $\la \tau|\tau\ra$ (and $\la w|w\ra$ ). Moreover, $\la \tau|\tau\ra$ itself can be quantized in different ways as
\begin{equation}
 \la \tau | \tau \ra= \bft_- \bftt    _+ - \bft_+ \bftt _- \dr 
\begin{cases} 
T_- \tT _+ - T_+ \tT _- = E \\
T_- \tT _+ - \tT _- T_+  = E+\one\\
\tT _+ T_-  - T_+ \tT _- = E+\one\\
\tT _+ T_-  - \tT _- T_+  = E+2\one,
\end{cases}
\end{equation}
where
\begin{equation}
E \ket{j,m} = 2j \ket{j,m}.
\end{equation}
For the term in the denominator, it is convenient to take one ordering choice which leads to $E+\one$ since it has also the effect to regularize the denominator when $j=0$. 
The sector in $w$ is quantized in an analogous way, but acts on covectors (bras), i.e.
\begin{equation}
\begin{aligned}
&|\tau \ra \dr \tT, \quad \,|\tau ] \dr T,\quad  \mbox{ both acting on } \ket{j,m} \\
 &|w \ra \dr \tT, \quad |w ] \dr T,\quad  \!\mbox{ both acting on } \bra{j,m} \\
&\la \tau|\tau\ra \dr E+\one, \quad   \la w|w\ra \dr E+\one,
\end{aligned}
\end{equation}
with
\begin{equation}
\bra{j,m} E = 2j \bra{ j,m}. 
\end{equation}
The matrix elements of the holonomy are easy to quantize, except for some ordering issues with the inverse of $E+1$. We have two natural quatization choices
\begin{equation}
g= \frac{-i}{(\la \tau|\tau\ra \la \taup|\taup\ra)^\demi} M \leadsto
\begin{cases}
M\dr \widehat M = \mat{cc}{
-\tT_-\ot \tT_++T_-\ot T_+& -T_-\ot T_-+\tT_-\ot \tT_-\\ 
-\tT_+\ot \tT_++T_+\ot T_+&-T_+\ot T_-+\tT_+\ot \tT_-}
\\[2em]
g \dr \widehat g =  \frac{-i}{\sqrt{E+1}} \widehat M \frac{1}{\sqrt{E+1}}
 \quad\mbox{ or }\quad  -i \widehat M  \frac{1}{{E+1}}.
\end{cases}
\end{equation} 
The first ordering choice (symmetric in $\sqrt{E+1}$) is in a way natural since when acting on the discrete series, we have that $\hat g_{11}= \hat g_{22}^\dagger$ and $\hat g_{12}= \hat g_{21}^\dagger$; note however than when acting on the continuous series, we only have that $\hat g_{11}= \hat g_{22}^t$ and $\hat g_{12}= \hat g_{21}^t$. The drawback of this ordering choice is that we cannot obtain from it the Biedenharn--Elliott relation, which is key to construct the quantum Hamiltonian constraint (cf section \ref{recover}). The second choice of ordering leads  to the Biedenharn--Elliott relation but the quantum matrix elements are not adjoint to each other unless one renormalizes\footnote{We thank E. Livine for discussions on this point.} the scalar product for the $\ket{j,m}$.


\section{Classical description of Lorentzian 3d LQG}
\begin{figure}[h]
\includegraphics[scale=.5]{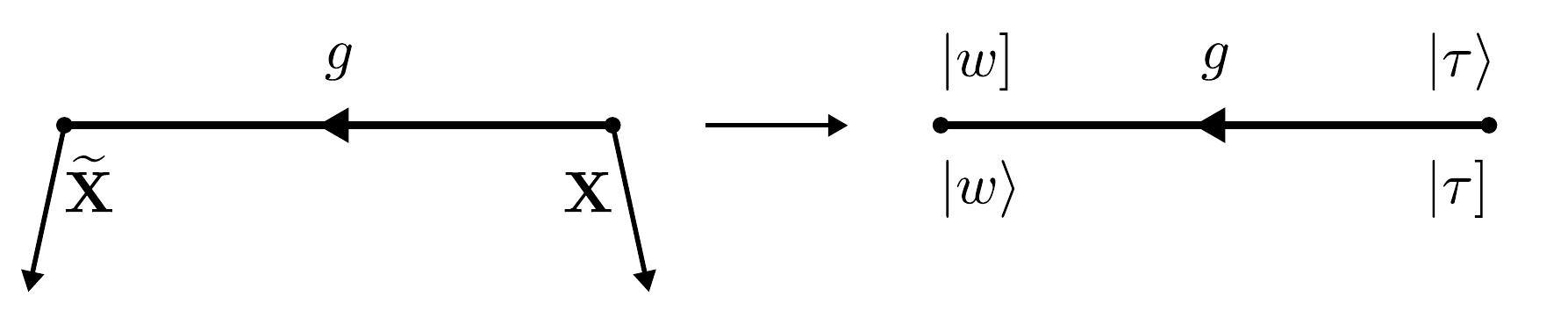} 
\caption{The information about fluxes is now encoded by a pair of spinors.}\label{fig1}
\end{figure}

We recall now the standard construction of the LQG phase space (see for example \cite{Bonzom:2011nv}), specializing it to the $\SU(1,1)$ case. The triad and connection $(e,\omega)$ are discretized into the flux and holonomy variables $(\bfX, g)\in T^*\SU(1,1)$. More precisely we consider a graph $\Gamma$ and looking at an edge $e$, we associate to the vertices of the edge the flux, whereas the edge is associated with the holonomy (cf Fig. \ref{fig1}). The flux $\widetilde \bfX$ is the flux parallel transported through the holonomy $g$.

The idea behind the spinorial framework is to  replace the fluxes and holonomies  by attaching a pair of spinors $|\tau\ra$, $|\tau]$ at each vertex. Given an edge, the two pairs of spinors provide the full information about $T^*\SU(1,1)$, since we can reconstruct from them the flux (cf \eqref{flux}) or the holonomy (cf \eqref{holo}). This is unlike the $(\bfX, \widetilde \bfX)$ variables which specify the holonomy only modulo a phase (a boost or a rotation according to the nature of $\bfX$ and $\widetilde \bfX$).

The dynamics of gravity is encoded by two constraints, the Gauss constraint and the flatness constraint. The Gauss constraint is discretized at the vertices of $\Gamma$. The discretized constraint corresponds to an (infinitesimal) $\SU(1,1)$ invariance at the vertex. Due to the proportionality between the fluxes and the $\su(1,1)$ generators, this invariance can be interpreted as saying that the total flux at each vertex is zero, i.e.
\be
\sum_i  \bfX_i = 0.
\ee  
One should be aware that this is merely an accident. When dealing with quantum groups, the invariance under the quantum group cannot be interpreted as saying that what plays the role of the fluxes sums up to zero\cite{Bonzom:2014bua}.

Given a vertex $v$, we can construct a set of functions which  commute with the Gauss constraint, and as such they will be called observables. They are defined in terms of  the spinors living on different legs of the vertex and    they are therefore $\SU(1,1)$ invariant. These functions are
\begin{equation}
\label{obs}
\begin{aligned}
f_{ab} &\equiv \cB(\bft_a,\bft_b) = \la\tau_a|\tau_b],
&\tf_{ab} &\equiv \cB(\bftt_a,\bftt_b) = [ \tau_a|\tau_b\ra,\\
e_{ab} &\equiv\cB(\bft_a,\bftt_b) = \la \tau_a|\tau_b\ra,
\quad &\te_{ab} &\equiv\cB(\bftt_a,\bft_b) = [ \tau_a|\tau_b]=-e_{ba}.
\end{aligned}
\end{equation}  
The observables $f_{ab}$ and $\tf_{ab}$  are not all independent when considering some reality conditions:  for example, if one uses either of the reality conditions $\bftt_{\pm}=-\ov \bft_\mp$ or $\bftt_{\pm}=\ov \bft_\mp$ on both of the legs $a$ and $b$, we get that $\tf_{ab}= \ov f_{ab}$. If instead we use $\bftt_{\pm}=-\ov \bft_\mp$ on leg $a$ and  $\bftt_{\pm}=\ov \bft_\mp$ on leg $b$ (or vice versa), we get $\tf_{ab}= -\ov f_{ab}$.

\begin{figure}[h]
\includegraphics[scale=1]{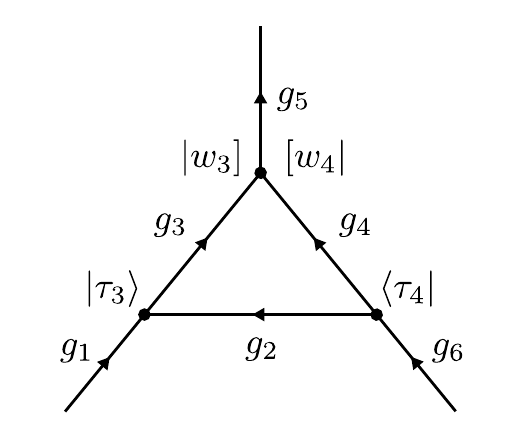}
\caption{The flatness constraint on the triangular face is $g_2 g_4\mone g_3=\one$.}\label{fig2} 
\end{figure}

The functions $e$, $\te$, $f$ and $\tf$ satisfy the closed Poisson relations
\begin{subequations}
\begin{align}
\{e_{ab}, e_{cd}\} &= -i(\delta _{cb}\,e_{ad} -\delta _{ad}\,e_{cb}), \\
\{e_{ab}, f_{cd}\} &= -i(\delta_{bc}\,f_{ad} - \delta_{bd}\,f_{ac}), \\
\{e_{ab}, \tf_{cd}\} &=- i(\delta_{ad}\,\tf_{bc}-\delta_{ac}\,\tf_{bd}), \\
\{f_{ab}, \tf_{cd}\}&=  -i(\delta _{cb}\,e_{ad} - \delta_{ad}\, \te_{cb}  - \delta_{ca} e_{bd} + \delta_{bd} \te_{ca}), \\
%
 \{f_{ab}, f_{cd}\}&= \{\tf_{ab}, \tf_{cd}\} =0.
\end{align}
\end{subequations}
As it is well-known now in the Euclidean case, we can use these observables to generate the standard  LQG observables, which are now expressed in terms of the observables associated to the intertwiners. 
The flatness constraint is discretized by asking that the product of the holonomies around each face $f$ of the graph is the identity, i.e.
\be\label{c1}
\prod_{e\in f} g_e = \one.
\ee
V. Bonzom and his collaborators  realized that this constraint (in the Euclidean case) can be recast in a natural constraint involving the fluxes \cite{Bonzom:2011hm} or the spinors \cite{Bonzom:2011nv}, according to the initial choice of variables on the graph. Essentially, one projects the flatness constraints on the basis provided by the spinors, to obtain a set of scalar constraints. The physical interpretation is that the scalar product of two spinors at a vertex is left invariant when parallel transporting the spinors (the fluxes) along the edges around the relevant face. Since one has different types of spinors, one  obtains different scalar constraints. Without loss of generality, we can focus on a triangular face of the graph, such as in Fig. \ref{fig2}. 
Sitting at the vertex between $g_2$ and $g_3$ and proceeding clockwise (i.e. along the cycle $\braket{342}$), the constraints are then 
\begin{equation}
\begin{aligned}
H^{[\ra}_{342}&:=
[\taup_2| (\one - g_2 g_4\mone g_3)|\tau_3\ra \la\taup_2|\tau_3],\quad
&
H^{\la\ra}_{342}&:=
\la\taup_2| (\one - g_2 g_4\mone g_3)|\tau_3\ra [\taup_2|\tau_3],\\
H^{\la ]}_{342}&:=
\la\taup_2| (\one - g_2 g_4\mone g_3)|\tau_3][\taup_2|\tau_3\ra,
&
H^{[]}_{342}&:=
[\taup_2| (\one - g_2 g_4\mone g_3)|\tau_3]\la\taup_2|\tau_3\ra.
\end{aligned}
\label{list}
\end{equation}

The constraint  \eqref{c1} is actually a set of 3 real scalar constraints: in fact $g_2 g_4\mone g_3$, as a $\SU(1,1)$ group element, is parametrized by 3 real parameters, so that the constraint 
\begin{equation}
\label{eq:342_constraint}
\one-g_2 g_4\mone g_3=0
\end{equation}
has 3 (real) degrees of freedom. The four complex constraints in \eqref{list}, being proportional to the matrix elements of \eqref{eq:342_constraint}, are equivalent to it and thus carry the same degrees of freedom.

Using the parallel transport of the spinors, we can simplify the expression of the previous Hamiltonian constraints. We can in particular express them in terms of the vertex observables $e_{ab}, f_{ab}, \widetilde f_{ab}$. For example, using \eqref{eq:g_action} for $\la w_2|g_2$ and $g_3|\tau_3]$, we have that 
\begin{equation}
\begin{split}
H^{\la ]}_{342}&=
\la\taup_2| (\one -  g_2 g_4\mone g_3 )|\tau_3] [\taup_2|\tau_3\ra\\
&=
 \left(\la\taup_2| \tau_3] - [ \tau_2| g_4\mone |\taup_3\ra\right)  [\taup_2|\tau_3\ra \\
&=
\left(\la\taup_2| \tau_3] -[\tau_2|  \left(i\frac{|\tau_4\ra [ \taup_4|- |\tau_4]\la \taup_4|}{\sqrt{\la \tau_4|\tau_4\ra}\sqrt{\la \taup_4|\taup_4\ra}}\right)       |\taup_3\ra \right)[\taup_2|\tau_3\ra \\
&=
\left( f_{23} -i\left(  \tf_{24} \tf_{43}-\te_{24} e_{43} \right)e_4\mone \right)\tf_{23} \\
&=
 f_{32}\tf_{32} -i\left(e_{43}\te_{24}  -\tf_{43}\tf_{24} \right)e_4\mone \tf_{32},
\label{hamiltonian constraint}
\end{split}
\end{equation}
where we used that $\la \tau_4|\tau_4\ra= \la \taup_4|\taup_4\ra = e_{44}=:e_4$, $f_{ab}= -f_{ba}$ and  $\tf_{ab} = -\tf_{ba} $.

Different sets of constraints can be obtained by considering the other possible cycles; the general expression for them is
\begin{equation}
\label{eq:general_hamiltonian_constraint}
\begin{aligned}
H_{abc}^{[\rangle}=& \widetilde{f}_{ac}f_{ac} + i\left( e_{ab}\widetilde{e}_{bc} - f_{ab}f_{bc} \vphantom{\tf}\right)e_b\mone f_{ac},\quad
&
H_{abc}^{\langle\rangle}=& \widetilde{e}_{ac}e_{ac} + i\left( e_{ab}\widetilde{f}_{bc} -f_{ab}e_{bc}\right)e_b\mone e_{ac},\\
H_{abc}^{\langle]}=& f_{ac}\widetilde{f}_{ac} - i\left( \widetilde{e}_{ab}e_{bc} - \widetilde{f}_{ab}\widetilde{f}_{bc} \right)e_b\mone \widetilde{f}_{ac},
&
H_{abc}^{[]}=& e_{ac}\widetilde{e}_{ac} - i\left( \widetilde{e}_{ab}f_{bc} -\widetilde{f}_{ab}\widetilde{e}_{bc}\right)e_b\mone \widetilde{e}_{ac},
\end{aligned}
\end{equation}
with
\begin{equation}
\braket{abc}=\braket{234},\braket{342},\braket{423}.
\end{equation}
The counter-clockwise cycles $\braket{432}$, $\braket{243}$ and $\braket{324}$ are ignored as they yield the same constraints as their clockwise counterparts.

One can check by direct computation that
\begin{equation}
H_{abc}^{[\rangle} + H_{abc}^{\langle]}-H_{abc}^{\langle\rangle}-H_{abc}^{[]}= e_a e_c \,\mathrm{tr}(\one - g_2 g_4\mone g_3),
\end{equation}
by making use of the property of the trace
\begin{equation}
\mathrm{tr}(ABC)=\mathrm{tr}(CAB)=\mathrm{tr}(BCA).
\end{equation}
\section{Relativistic spin networks}
We would like to quantize the classical LQG data we have just described.  In this section, we  construct the kinematical  Hilbert, given in terms of $\SU(1,1)$ spin networks.  L. Freidel and E. Livine have studied in details spin networks defined for a non-compact group \cite{non-compact}. Due to the non-compacticity of the group, divergencies might arise, but they prescribed a way to deal with them. They used spin networks represented as functions over the group, whereas  we here are interested in spin networks defined in terms of representations (the dual picture);  we review therefore  the $\SU(1,1)$ spin networks from the representation point of view.

The $\SU(1,1)$ recoupling theory has been recalled in section \ref{recoupling_theory}.  We mainly focus  on the fundamental building block of a spin network, namely the 3-valent intertwiner. We also discuss the 4-valent intertwiner, as it allows to introduce the Racah coefficient. Finally, we recall  the scalar product we can use in our framework, to equip the vector space of spin networks with a Hilbert space structure.

\subsection{Intertwiners}
Intertwiners are the morphisms in the category of (linear) group representations, i.e. structure-preserving maps from one representation to another. They are defined as follows: an intertwiner between two \emph{arbitrary} representations
\begin{equation}
\rho:\SLTR\rightarrow \mathrm{GL}(V),\quad \pi:\SLTR\rightarrow \mathrm{GL}(W),
\end{equation}
not necessarily irreducible, is a linear map $\psi:V\rightarrow W$ such that
\begin{equation}
\psi\circ \rho(g)=\pi(g)\circ \psi,\quad \forall g\in\SLTR.
\end{equation}
The set of all possible intertwiners from $\rho$ to $\pi$ forms a vector space, which will be denoted by $\Hom_G(\rho,\pi)$, where $G:=\SU(1,1)$.

We will only work with intertwiners between representations that are either irreducible or a tensor product of irreducible ones\footnote{When speaking of products we assume none of the representations involved is the trivial one, for obvious reasons.}. We will call an intertwiner
\begin{equation}
\psi:\bigotimes_{a=1}^k V_{j_a}\rightarrow \bigotimes_{b=r+1}^n V_{j_b}
\end{equation}
an $n$-valent intertwiner and, for reasons that will become clear shortly, we will say it has $k$ incoming legs and $(n-k)$ outgoing ones.

Of particular interest are the $3$-valent intertwiners. The vector space $\Hom_G(\rho_{j_1}\otimes \rho_{j_2},\rho_{j_3})$, if the decomposition of $\rho_{j_1}\otimes \rho_{j_2}$ exists, is completely specified by this decomposition. In fact, one can show that a non-vanishing intertwiner only exists if $\rho_{j_3}$ appears in the decomposition; the number of independent intertwiners equals the multiplicity of $\rho_{j_3}$ in the decomposition (1 or 2 for the known decompositions). These basis elements will be denoted by
\begin{equation}
\spinnet{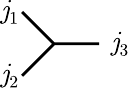}:\ket{j_1,m_1}\otimes\ket{j_2,m_2}\in V_{j_1}\otimes V_{j_2}\mapsto \sum_{m_3\in \mathcal{M}_{j_3}}B(j_3,m_3|j_1,m_1;j_2,m_2)\ket{j_3,m_3}\in V_{j_3},
\end{equation}
where we assume $j_3$ also includes an appropriate label for multiplicities, when necessary. On the l.h.s. we used a graphical notation for the map, which will turn out to be very useful. It is to be read this way: incoming representations (legs) are on the left, while outgoing ones are on the right; an arrow will be used to make the direction clear if needed.

Analogously, the basis elements for $\Hom_G(\rho_{j_3},\rho_{j_1}\otimes \rho_{j_2})$ are given by the intertwiners
\begin{equation}
\spinnet{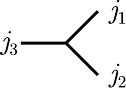}:\ket{j_3,m_3}\in V_{j_3}\mapsto \sum_{\substack{m_1\in\mathcal{M}_{j_1} \\ m_2 \in \mathcal{M}_{j_2}}}A(j_1,m_1;j_2,m_2|j_3,m_3)\ket{j_1,m_1}\otimes\ket{j_2,m_2}\in V_{j_1}\otimes V_{j_2}.
\end{equation}
Moreover, we will denote the unique intertwiner in the $1$-dimensional space $\Hom_G(\rho_j,\rho_j)$ by
\begin{equation}
\spinnet{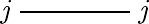}\equiv \mathbf{id}_{V_j}.
\end{equation}
The two kinds of $3$-valent intertwiners can be used as building blocks of all the others, provided that the necessary Clebsch--Gordan decomposition exist. This can be achieved by composing intertwiners, to obtain maps on bigger representations; with our graphical notation, this amount to ``glueing'' them together.

We will call any such composition of intertwiners a spin network.
Note that, when working with unitary representations, there is no way to obtain a \emph{closed} spin network\footnote{i.e. an element of $\Hom_G(F_0,F_0)$.} with this glueing procedure, as the trivial representation $F_0$ does not appear in any recoupling of infinite-dimensional representations. Closing a spin network by tracing, which graphically amounts to connecting an incoming leg with an incoming one of the same intertwiner, leads to divergencies, which will have to be dealt with. In this article however we are only interested in the nodes \emph{inside} a spin network, which would be unaffected by any regularization procedure.

\subsection{Racah coefficients}\label{sec:racah}

Consider a $4$-valent intertwiner $\psi$ with 3 incoming legs $\rho_{j_1}\otimes \rho_{j_2}\otimes \rho_{j_3}$ and a single outgoing one $\rho_{j}$: it will generally not be unique, unless one of the representations involved is the trivial one. Two possible \emph{bases} of intertwiners, whose linear combinations can be used to construct any $4$-valent one of this type can be obtained by exploiting the associativity of tensor products, i.e.
\begin{equation}
V_{j_1}\otimes V_{j_2}\otimes V_{j_3}\cong (V_{j_1}\otimes V_{j_2})\otimes V_{j_3} \cong V_{j_1}\otimes (V_{j_2}\otimes V_{j_3}).
\end{equation}
Assuming the decomposition in irreducible representations of $V_{j_1}\otimes V_{j_2}$ exists, the generic $\psi$ can be written as a linear combinations of the intertwiners
\begin{equation}\label{eq:4valent_basis1}
\spinnet{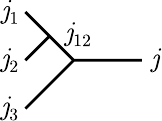},\quad j_{12}\in\mathcal{D}(j_1,j_2);
\end{equation}
analogously, if $V_{j_2}\otimes V_{j_3}$ is decomposable, the intertwiners
\begin{equation}\label{eq:4valent_basis2}
\spinnet{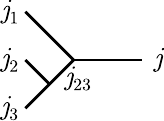},\quad j_{23}\in\mathcal{D}(j_2,j_3)
\end{equation}
form a basis as well. We will now study how the two bases are related to each other.

First notice that\footnote{To simplify notation, the range of the $j$'s in the summation is omitted: it is implied to only assume the values for which a non-vanishing intertwiner exists. Moreover, when a subset of labels $j$ appears continuously in a decomposition, over that subset the sum is to be considered an integration.} 
\begin{equation}
\sum_{j_{12},j}\spinnet{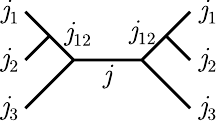}=\sum_{j_{23},j}\spinnet{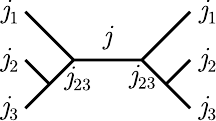}=\spinnet{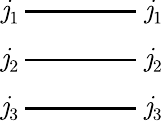},
\end{equation}
as can be checked explicitly using the properties of Clebsch--Gordan coefficients. This equation can be ``glued'' to the basis elements \eqref{eq:4valent_basis1} to obtain
\begin{equation}
\spinnet{racah/j12}=\sum_{j_{23},j'}\spinnet{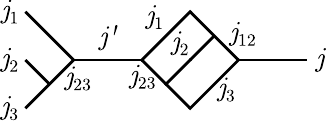}.
\end{equation}
Now, the intertwiner
\begin{equation}
\spinnet{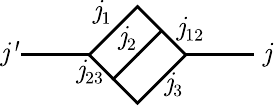},
\end{equation}
having only one incoming and outgoing representation, must \emph{necessarily} be proportional to the unique intertwiner between $j'$ and $j$. Since the latter vanishes when $j\neq j'$, it must be
\begin{equation}\label{eq:racah1}
\spinnet{racah/racah_int1}\propto
\delta(j,j')\spinnet{id/j-j},
\end{equation}
where the $\delta$ is to be considered a Dirac delta over continuous subsets in both $\mathcal{D}(j_{12},j_3)$ and $\mathcal{D}(j_1,j_{23})$, and a Kronecker delta otherwise. The proportionality factor in \eqref{eq:racah1}, which we will call \emph{Racah coefficient}, is given by\footnote{To make the equation more readable the $j$'s were dropped. Which $j$ goes where can be easily inferred by the subscripts on the $m$'s: for example $A(m_1;m_{23}|m)\equiv A(j_1,m_1;j_{23},m_{23}|j,m)$.}
\begin{equation}\label{eq:racah2}
\begin{bmatrix}
j_1 & j_2 & j_{12}\\
j_3 & j & j_{23}
\end{bmatrix}=
\sum_{\substack{m_1,m_2,m_3\\m_{12},m_{23}}}
A(m_1;m_{23}|m)A(m_2;m_3|m_{23})
B(m_{12}|m_1;m_2)B(m|m_{12};m_3),
\end{equation}
with $m\in\mathcal{M}_j$; one can check using the Clebsch--Gordan recursion relations that the result does not depend on which $m$ one chooses. We finally get that
\begin{equation}
\label{eq:racah_coeff1}
\spinnet{racah/j12}
=\sum_{j_{23}}
\begin{bmatrix}
j_1 & j_2 & j_{12}\\
j_3 & j   & j_{23}
\end{bmatrix}
\spinnet{racah/j23},
\end{equation}
i.e. the Racah coefficients are the components of the elements of one basis in terms of the other. An analogous argument can be made for the basis elements \eqref{eq:4valent_basis2}. With our convention for the Clebsch--Gordan coefficients, one can check that
\begin{equation}
\spinnet{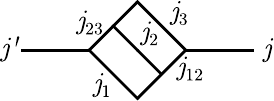}=\spinnet{racah/racah_int1}
\end{equation}
so that
\begin{equation}
\label{eq:racah_coeff2}
\spinnet{racah/j23}
=\sum_{j_{12}}
\begin{bmatrix}
j_1 & j_2 & j_{12}\\
j_3 & j & j_{23}
\end{bmatrix}
\spinnet{racah/j12}.
\end{equation}

We stress there was \textit{no} mention of unitary representations in our discussion of Racah coefficients. What we presented is well-defined any time the appropriate Clebsch--Gordan decomposition exist. This means in particular that we can consider Racah coefficients \textit{involving both unitary and non-unitary representations}. This will be relevant when discussing the quantum version of the observables $e$, $f$ and $\widetilde f$.

\subsection{Hilbert space structure}

We already mentioned that $\Hom_G(\rho,\pi)$ is a vector space; we will show now how an inner product can be defined naturally on it. This time we will also require that the irreducible representations are only the ones appearing in the Plancherel decomposition.

The space of intertwiners can inherit an inner product space by requiring that\footnote{Note that it is always true that the l.h.s can be split in the sum of independent subspaces on the right: what we are really requiring is for these subspaces to be orthogonal.}
\begin{equation}
\Hom_G(\textstyle \bigotimes_a\rho_{j_a}\oplus\bigotimes_b \rho_{j_b},\bigotimes_c \rho_{j_c})\equiv
\Hom_G(\textstyle \bigotimes_a\rho_{j_a},\bigotimes_c \rho_{j_c}) \oplus
\Hom_G(\textstyle \bigotimes_b \rho_{j_b},\bigotimes_c \rho_{j_c}).
\end{equation}
It is then easy to convince oneself that the composition of two intertwiners belongs to (a space isomorphic to) the tensor product of their respective intertwiner spaces, so that, for example,
\begin{equation}
\Hom_G(\rho_{j_1}\otimes\rho_{j_2},\rho_{j_3}\otimes \rho_{j_4})=
\int^\oplus \eder\xi(j)\Hom_G(\rho_{j_1}\otimes\rho_{j_2},\rho_j)\otimes\Hom_G(\rho_{j},\rho_{j_3}\otimes \rho_{j_4})
\end{equation}
or
\begin{equation}
\Hom_G(\rho_{j_1}\otimes\rho_{j_2}\otimes\rho_{j_3},\rho_{j})=
\int^\oplus \eder\xi(j_{12})\Hom_G(\rho_{j_1}\otimes\rho_{j_2},\rho_{j_{12}})\otimes\Hom_G(\rho_{j_{12}}\otimes\rho_{j_3},\rho_j).
\end{equation}
We can repeat this process until we only have sums of products of $3$-valent spaces, so that it only remains to define the inner product on the latter. This is easily achieved:
\begin{itemize}
\item when the space is one dimensional there is only one basis vector which we may normalize to $1$;
\item when the space is two dimensional, i.e. there is multiplicity, we choose the two basis elements to be orthonormal.
\end{itemize}
One can check explicitly that this is consistent with the possibility of using different decompositions for the same space, e.g.
\begin{equation}
\Hom_G(\rho_{j_1}\otimes\rho_{j_2}\otimes\rho_{j_3},\rho_{j})=
\int^\oplus \eder\xi(j_{23})\Hom_G(\rho_{j_2}\otimes\rho_{j_3},\rho_{j_{23}})\otimes\Hom_G(\rho_{j_1}\otimes \rho_{j_{23}},\rho_{j}),
\end{equation}
so that the procedure is well defined.

Restricting ourselves to representations in the Plancherel decomposition makes our construction possible, as it guarantees that the direct sums in the Clebsch--Gordan decomposition are orthogonal. However, one should note that if, instead, we only use finite-dimensional representations, the same will be true and the inner product will still be well defined.

\section{3D Lorentzian LQG and  Lorentzian Ponzano--Regge model}
In this section, we first construct the quantum version of the spinor observables and determine their action on intertwiners. We highlight the fact that, in some cases, they might map unitary representations to non-unitary ones. We then discuss some properties of the Racah coefficients, which are defined when involving both unitary or non-unitary representations. We pinpoint in particular that the Biedenharn--Elliott relation is very general and holds even when involving some finite-dimensional (hence non-unitary) representations. Finally we show how the quantum version of the Hamiltonian  constraint has in its kernel the Racah coefficient, which generates the Lorentzian Ponzano--Regge model. The proof consists in showing that the quantum Hamiltonian constraint implements a recursion relation, essentially the Biedenharn--Elliott,  whose solution is the Racah coefficient.  
\subsection{Intertwiner observables} \label{sec:int_obs}\label{lqg-obs}
We want observables in LQG to be invariant under the action of the gauge group $\SLTR$: this is exactly what \emph{scalar operators} (tensor operators of rank $0$) are.
The usual observables we consider are those built from the algebra generators, which are essentially the components of the vector operator $V$ defined in \eqref{eq:vector operator}. When acting on a product of representations $\bigotimes_a \rho_{j_a}$ they are defined as
\begin{equation}
Q_{ab}:=\frac{\sqrt{3}}{2}\sum_{\mu}A(1,\mu;1,-\mu|0,0)\,V^a_\mu V^b_{-\mu}=-J^a_0J^b_0 +\frac{1}{2}\left(J^a_- J^b_+ + J^a_+ J^b_-\right),
\end{equation}
where $V^a$ denotes the operator acting only on representation $a$. Defining
\begin{equation}
J_1:=\frac{1}{2}\left(J_+ + J_- \right)\equiv -iX_1,\quad J_2:=\frac{1}{2i}\left(J_+ - J_- \right)\equiv -iX_2,
\end{equation}
we get the suggestive equality
\begin{equation}
Q_{ab}=\eta^{ij}J^a_i J^b_j,\quad \eta=\mathrm{diag}(-1,1,1),
\end{equation}
where $i$ and $j$ are space-time indices.

When working in the spinorial setting, we can construct  scalar operators by combining the two spinor operators $T$ and $\widetilde T$. The four  kinds of operators we can get are
\begin{subequations}
\begin{align}
E_{ab}&=-\sqrt{2}\sum_{\mu}A(\tfrac{1}{2},\mu;\tfrac{1}{2},-\mu|0,0)\,T^a_\mu \widetilde{T}^b_{-\mu}= T^a_- \widetilde{T}^b_+ - T^a_+ \widetilde{T}^b_-,
\\
F_{ab}&=-\sqrt{2}\sum_{\mu}A(\tfrac{1}{2},\mu;\tfrac{1}{2},-\mu|0,0)\,T^a_\mu {T}^b_{-\mu}= T^a_- {T}^b_+ - T^a_+ {T}^b_- ,
\\
\widetilde{F}_{ab}&=-\sqrt{2}\sum_{\mu}A(\tfrac{1}{2},\mu;\tfrac{1}{2},-\mu|0,0)\,\widetilde{T}^a_\mu \widetilde{T}^b_{-\mu}= \widetilde{T}^a_- \widetilde{T}^b_+ - \widetilde{T}^a_+ \widetilde{T}^b_-,\\
\widetilde{E}_{ab}&=-\sqrt{2}\sum_{\mu}A(\tfrac{1}{2},\mu;\tfrac{1}{2},-\mu|0,0)\,\widetilde{T}^a_\mu {T}^b_{-\mu}=\widetilde{T}^a_- T^b_+ - \widetilde{T}^a_+ T^b_-\equiv-E_{ba}-2\delta_{ab}\mathbbm{1}.
\end{align}
\end{subequations}
These operators are the quantum analogues of the classical observables \eqref{obs}; one should note, however, that an ordering factor appears in the quantization of $\widetilde e_{ab}$. Just as in the classical case, they form a closed operator algebra, with commutation relations
\begin{subequations}
\begin{align}
[E_{ab},E_{cd}]&=\delta_{cb}E_{ad}-\delta_{ad}E_{cb},\\
[E_{ab},F_{cd}]&=\delta_{bc}F_{ad}-\delta_{bd}F_{ac},\\
[E_{ab},\widetilde{F}_{cd}]&=\delta_{ad}\widetilde{F}_{bc}-\delta_{ac}\widetilde{F}_{bd},\\
[F_{ab},\widetilde{F}_{cd}]&=\delta_{cb}E_{ad}-\delta_{ad}\widetilde{E}_{cb}-\delta_{ca}E_{bd}+\delta_{bd}\widetilde{E}_{ca},\\
[F_{ab},F_{cd}]&=[\widetilde{F}_{ab},\widetilde{F}_{cd}]=0.
\end{align}
\end{subequations}
Note that, when acting on the continuous class, the operators $E,F,\widetilde F$  take unitary representations (in the Plancherel decomposition) to a \emph{non-unitary} ones\footnote{The $T$ operators would send $-\tfrac{1}{2}+is$ to $is$ while $\widetilde T$ would sent it to $-1+is$, both of which would not be unitary anymore.}.
As such, they are not proper observables when acting on continuous representations; however, one can choose quadratic functions of these observables such that the representation is sent to itself. 

Due to the relation between $T$ and $\widetilde{T}$, the operators we defined are not all independent. One has, in general,
\begin{equation}
F^\transpose_{ab}=\widetilde{F}_{ab},\quad E^\transpose_{ab}=-\widetilde{E}_{ab}.
\end{equation}
In particular cases the transposes can be converted to adjoints; for example, if $a$ and $b$ both denote a representation in the discrete positive (negative) class $F^\transpose_{ab}=F^\dagger_{ab}$, while if one is them is discrete positive and the other discrete negative $F^\transpose_{ab}=-F^\dagger_{ab}$.
 
The operators we defined act on representations; their action can be extended to intertwiners as follows. Let
\begin{equation}
\psi:\bigotimes_{a=1}^k\ket{j_a,m_a}\rightarrow \sum_{m_{r+1}}\dotsm\sum_{m_n}\alpha(m_1,\dotsc,m_n)\bigotimes_{b=k+1}^n\ket{j_b,m_b}
\end{equation}
be a generic $n$-valent intertwiner\footnote{The operators we define will always act on a single \emph{node}, i.e. $n$-valent intertwiner, inside a generic spin network. }, where $\alpha$ is a function depending on Clebsch--Gordan coefficients. The condition for $\psi$ to be an intertwiner is translated in terms of this functions as
\begin{equation}\label{eq:intertwiner_conditions}
\begin{cases}
\alpha(m_1,\dotsc,m_n)\sum_{a=1}^k m_a =\alpha(m_1,\dotsc,m_n) \sum_{b=k+1}^n m_b
\\[0.7em]
\sum_{c=1}^k C_\pm(j_c,m_c)\alpha(\vec{m}^{(in)}\pm\boldsymbol{\delta}^{(\text{in})}_{c},\vec{m}^{(\text{out})})=\sum_{d=k+1}^n C_\mp(j_d,m_d)\alpha(\vec{m}^{(\text{in})},\vec{m}^{(\text{out})} \mp\boldsymbol{\delta}^{(\text{out})}_{d}),
\end{cases}
\end{equation}
where
\begin{equation}
\vec{m}^{(\text{in})}:=(m_1,\dotsc,m_k),\quad \vec{m}^{(\text{out})}:=(m_{k+1},\dotsc,m_n)
\end{equation}
and
\begin{equation}
\boldsymbol{\delta}^{(\text{in})}_{c}:=(\delta_{1c},\dotsc,\delta_{kc}),\quad \boldsymbol{\delta}^{(\text{out})}_{d}:=(\delta_{k+1,d},\dotsc,\delta_{nd}).
\end{equation}
The intertwiner $\psi$ can be also expressed in the form
\begin{equation}\label{eq:dualized_intertwiner}
\psi^\star :=\sum_{m_1}\dotsm\sum_{m_n}\alpha(m_1,\dotsc,m_n)\bigotimes_{b=k+1}^n\ket{j_b,m_b}\otimes \bigotimes_{a=1}^k\bra{j_a,m_a},
\end{equation}
which does indeed return the same values when acting on $\bigotimes_a V_{j_a}$. However, \eqref{eq:dualized_intertwiner} is not generally an element of the Hilbert space $\bigotimes_b V_{J_b}\otimes \bigotimes_a V^*_{j_a}$, since it does not generally have finite norm for infinite-dimensional representations\footnote{This can be easily checked in the case of an intertwiner with one incoming and one outgoing leg, which is necessarily proportional to the identity.}: it is only to be regarded as a formal expression, similarly to the usual way of representing the identity of a separable Hilbert space as
\begin{equation}
\sum_{i\in I}\ket{i}\bra{i},
\end{equation}
where $\{\ket{i}\}_{i\in I}$ is an orthonormal basis.

One can easily check, using \eqref{eq:intertwiner_conditions}, that $\psi$ is in intertwiner if and only if
\begin{equation}
J_0\psi^\star=0,\quad J_\pm\psi^\star=0,
\end{equation}
where the generators act on dual vectors as the dual representation (see Appendix \ref{app:dual_rep}), i.e.
\begin{equation}
J_0\bra{j,m}=-m\bra{j,m},\quad J_\pm\bra{j,m}=-C_\mp(j,m)\bra{j,m\mp 1}.
\end{equation}
The action of an operator of the form
\begin{equation}\label{eq:int_operator}
T:\bigotimes_{b=k+1}^n V_{j_b}\otimes \bigotimes_{a=1}^k V^*_{j_a}\rightarrow \bigotimes_{b=k+1}^n V_{j'_b}\otimes \bigotimes_{a=1}^k V^*_{j'_a}
\end{equation}
is then defined by inverting transformation \eqref{eq:dualized_intertwiner} for $T\psi^\star$. One can check that  the resulting map is an intertwiner if and only if $T$ is a scalar operator.

The $E$, $F$ and $\widetilde{F}$ operators can be expressed as a sum of operators of the form \eqref{eq:int_operator} by having $T^a$ and $\widetilde{T}^a$ act as the identity on anything but the $a$-th leg (incoming or outgoing) and by extending their action to dual vectors. This can be achieved by making use of the isomorphism of representations (see Appendix \ref{app:dual_rep})
\begin{equation}
\varphi_j:\bra{j,m}\mapsto (-1)^m\ket{j,-m};
\end{equation}
explicitly, we define
\begin{equation}
T_\pm\bra{j,m}:=\varphi^{-1}_{j+\frac{1}{2}}T_\pm\varphi_j\bra{j,m},\quad \widetilde{T}_\pm\bra{j,m}:=\varphi^{-1}_{j-\frac{1}{2}}\widetilde{T}_\pm\varphi_j\bra{j,m}.
\end{equation}
We list here the action of the scalar operators on some $3$-valent intertwiners of interest, where we make use of the notation
\begin{equation}
D(j):=\sqrt{2j+1};
\end{equation}
these actions are\footnote{Note that, by definition, $F_{ba}=-F_{ab}$ and $\widetilde{F}_{ba}=-\widetilde{F}_{ab}$.}
\begin{subequations}
\begin{align}
E_{12}\spinnet{trivalent/j1xj2-j3}&=D(j_1)D(k_2)
\racah{j_1}{\tfrac{1}{2}}{k_1}{k_2}{j_3}{j_2}
\delta_{k_1,j_1+\frac{1}{2}}\delta_{k_2,j_2-\frac{1}{2}}
\spinnet{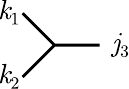},
\\[0.7em]
E_{21}\spinnet{trivalent/j1xj2-j3}&=-D(j_1)D(k_2)
\racah{j_1}{\tfrac{1}{2}}{k_1}{k_2}{j_3}{j_2}
\delta_{k_1,j_1-\frac{1}{2}}\delta_{k_2,j_2+\frac{1}{2}}
\spinnet{trivalent/k1xk2-j3},
\\[0.7em]
F_{12}\spinnet{trivalent/j1xj2-j3}&=D(j_1)D(k_2)
\racah{j_1}{\tfrac{1}{2}}{k_1}{k_2}{j_3}{j_2}
\delta_{k_1,j_1+\frac{1}{2}}\delta_{k_2,j_2+\frac{1}{2}}
\spinnet{trivalent/k1xk2-j3},
\\[0.7em]
\widetilde{F}_{12}\spinnet{trivalent/j1xj2-j3}&=D(j_1)D(k_2)
\racah{j_1}{\tfrac{1}{2}}{k_1}{k_2}{j_3}{j_2}
\delta_{k_1,j_1-\frac{1}{2}}\delta_{k_2,j_2-\frac{1}{2}}
\spinnet{trivalent/k1xk2-j3},
\end{align}
\end{subequations}
\begin{subequations}
\begin{align}
E_{12}\spinnet{trivalent/j3-j1xj2}&=D(j_1)D(k_2)
\racah{j_1}{\tfrac{1}{2}}{k_1}{k_2}{j_3}{j_2}
\delta_{k_1,j_1+\frac{1}{2}}\delta_{k_2,j_2-\frac{1}{2}}
\spinnet{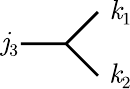},
\\[0.7em]
E_{21}\spinnet{trivalent/j3-j1xj2}&=D(j_1)D(k_2)
\racah{j_1}{\tfrac{1}{2}}{k_1}{k_2}{j_3}{j_2}
\delta_{k_1,j_1-\frac{1}{2}}\delta_{k_2,j_2+\frac{1}{2}}
\spinnet{trivalent/j3-k1xk2},
\\[0.7em]
F_{12}\spinnet{trivalent/j3-j1xj2}&=-D(j_1)D(k_2)
\racah{j_1}{\tfrac{1}{2}}{k_1}{k_2}{j_3}{j_2}
\delta_{k_1,j_1+\frac{1}{2}}\delta_{k_2,j_2+\frac{1}{2}}
\spinnet{trivalent/j3-k1xk2},
\\[0.7em]
\widetilde{F}_{12}\spinnet{trivalent/j3-j1xj2}&=-D(j_1)D(k_2)
\racah{j_1}{\tfrac{1}{2}}{k_1}{k_2}{j_3}{j_2}
\delta_{k_1,j_1-\frac{1}{2}}\delta_{k_2,j_2-\frac{1}{2}}
\spinnet{trivalent/j3-k1xk2},
\end{align}
\end{subequations}
\begin{subequations}
\begin{align}
E_{23}\spinnet{trivalent/j1xj2-j3}&=i D(j_2)D(j_3)
\racah{j_1}{j_2}{j_3}{\tfrac{1}{2}}{k_3}{k_2}
\delta_{k_2,j_2+\frac{1}{2}}\delta_{k_3,j_3-\frac{1}{2}}
\spinnet{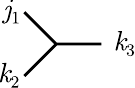},
\\[0.7em]
E_{32}\spinnet{trivalent/j1xj2-j3}&=i D(j_2)D(j_3)
\racah{j_1}{j_2}{j_3}{\tfrac{1}{2}}{k_3}{k_2}
\delta_{k_2,j_2-\frac{1}{2}}\delta_{k_3,j_3+\frac{1}{2}}
\spinnet{trivalent/j1xk2-k3},
\\[0.7em]
F_{23}\spinnet{trivalent/j1xj2-j3}&=-i D(j_2)D(j_3)
\racah{j_1}{j_2}{j_3}{\tfrac{1}{2}}{k_3}{k_2}
\delta_{k_2,j_2+\frac{1}{2}}\delta_{k_3,j_3+\frac{1}{2}}
\spinnet{trivalent/j1xk2-k3},
\\[0.7em]
\widetilde{F}_{23}\spinnet{trivalent/j1xj2-j3}&=i D(j_2)D(j_3)
\racah{j_1}{j_2}{j_3}{\tfrac{1}{2}}{k_3}{k_2}
\delta_{k_2,j_2-\frac{1}{2}}\delta_{k_3,j_3-\frac{1}{2}}
\spinnet{trivalent/j1xk2-k3}.
\end{align}
\end{subequations}

The Racah coefficients we have used may involve both unitary and non-unitary representations; as we discussed in section \ref{sec:racah}, they are still well defined in this case.

\subsection{Biedenharn--Elliott relation and symmetries of the Racah coefficients}

\subsubsection{Symmetries}

In order to proceed with our calculations, we are going to need some properties of the Racah coefficients, valid when at least one of $\rho_{j_1}$, $\rho_{j_2}$ and $\rho_{j_3}$ is the finite-dimensional representation $F_\frac{1}{2}$. The proof of these symmetries is straightforward, and can be checked by inserting explicitly the Clebsch--Gordan coefficients from Table \ref{CG-finite} in \eqref{eq:racah2}. These symmetries are
\begin{subequations}
\begin{align}
\begin{bmatrix}
j_1 & \tfrac{1}{2} & k_1\\
j_2 & J & k_2 
\end{bmatrix}
&= (-1)^{j_1+j_2-k_1-k_2}\frac{D(k_1)D(k_2)}{D(j_1)D(j_2)}
\begin{bmatrix}
k_1 & \tfrac{1}{2} & j_1\\
k_2 & J & j_2 
\end{bmatrix},
\\[0.7em]
\begin{bmatrix}
\tfrac{1}{2} & j_1& k_1\\
J & j_2 & k_2 
\end{bmatrix}
&= (-1)^{j_1+j_2-k_1-k_2}\frac{D(k_1)D(j_2)}{D(j_1)D(k_2)}
\begin{bmatrix}
\tfrac{1}{2} & k_1& j_1\\
J & k_2 & j_2 
\end{bmatrix},
\\[0.7em]
\begin{bmatrix}
J & j_1 & j_2 \\
\tfrac{1}{2} & k_2& k_1
\end{bmatrix}
&= (-1)^{j_1+k_2-k_1-j_2}\frac{D(k_1)D(k_2)}{D(j_1)D(j_2)}
\begin{bmatrix}
J & k_1 & k_2 \\
\tfrac{1}{2} & j_2& j_1
\end{bmatrix}.
\end{align}
\end{subequations}
Note that the numbers on the exponents are always in $\mathbb{Z}/2$: for example, in the first equation, it must be $k_i\in\mathcal{D}(\frac{1}{2},j_i)$ so that $j_i-k_i=\pm\frac{1}{2}$.

\subsubsection{Biedenharn--Elliott}

Racah coefficients, \textit{regardless} of the representation classes involved, satisfy the so-called \emph{Biedenharn--Elliott relation} or \emph{pentagon identity}. 
\begin{figure}
\includegraphics[scale=.7]{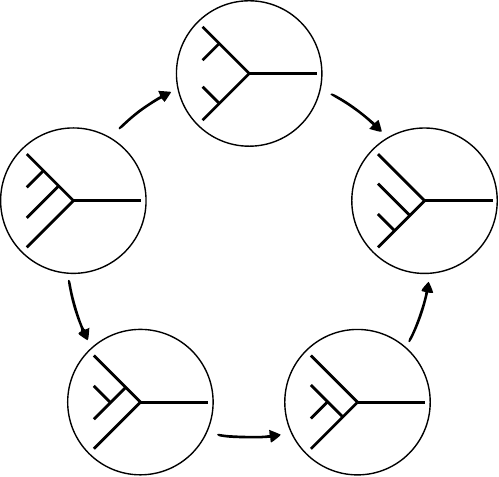}
\caption{The pentagon identity.}\label{fig:pentagon}
\end{figure}
This can be represented graphically as in Figure \ref{fig:pentagon}: one can go from the leftmost representation to the rightmost one by repeated Racah transformations in two possible ways. By equating the Racah coefficients appearing in the two transformations one gets that
\begin{subequations}
\begin{gather}
\sum_{j_{23}}
\begin{bmatrix}
j_1 & j_2 & j_{12}\\
j_3 & j_{123} & j_{23}
\end{bmatrix}
\begin{bmatrix}
j_1 & j_{23} & j_{123}\\
j_4 & j & j_{234}
\end{bmatrix}
\begin{bmatrix}
j_2 & j_3 & j_{23}\\
j_4 & j_{234} & j_{34}
\end{bmatrix}
=
\begin{bmatrix}
j_1 & j_2 & j_{12}\\
j_{34} & j & j_{234}
\end{bmatrix}
\begin{bmatrix}
j_{12} & j_3 & j_{123}\\
j_4 & j & j_{34}
\end{bmatrix}.
\intertext{Analogously, one can repeat the process starting from one of the other 4 intertwiners, to get the remaining identities}
\sum_{j_{123}}
\begin{bmatrix}
j_{12} & j_3 & j_{123}\\
j_4 & j & j_{34}
\end{bmatrix}
\begin{bmatrix}
j_1 & j_2 & j_{12}\\
j_3 & j_{123} & j_{23}
\end{bmatrix}
\begin{bmatrix}
j_1 & j_{23} & j_{123}\\
j_4 & j & j_{234}
\end{bmatrix}
=
\begin{bmatrix}
j_2 & j_3 & j_{23}\\
j_4 & j_{234} & j_{34}
\end{bmatrix}
\begin{bmatrix}
j_1 & j_2 & j_{12}\\
j_{34} & j & j_{234}
\end{bmatrix},
\\[0.7em]
\sum_{j_{12}}
\begin{bmatrix}
j_1 & j_2 & j_{12}\\
j_{34} & j & j_{234}
\end{bmatrix}
\begin{bmatrix}
j_{12} & j_3 & j_{123}\\
j_4 & j & j_{34}
\end{bmatrix}
\begin{bmatrix}
j_1 & j_2 & j_{12}\\
j_3 & j_{123} & j_{23}
\end{bmatrix}
=
\begin{bmatrix}
j_1 & j_{23} & j_{123}\\
j_4 & j & j_{234}
\end{bmatrix}
\begin{bmatrix}
j_2 & j_3 & j_{23}\\
j_4 & j_{234} & j_{34}
\end{bmatrix},
\\[0.7em]
\sum_{j_{34}}
\begin{bmatrix}
j_2 & j_3 & j_{23}\\
j_4 & j_{234} & j_{34}
\end{bmatrix}
\begin{bmatrix}
j_1 & j_2 & j_{12}\\
j_{34} & j & j_{234}
\end{bmatrix}
\begin{bmatrix}
j_{12} & j_3 & j_{123}\\
j_4 & j & j_{34}
\end{bmatrix}
=
\begin{bmatrix}
j_1 & j_2 & j_{12}\\
j_3 & j_{123} & j_{23}
\end{bmatrix}
\begin{bmatrix}
j_1 & j_{23} & j_{123}\\
j_4 & j & j_{234}
\end{bmatrix},
\\[0.7em]
\sum_{j_{234}}
\begin{bmatrix}
j_1 & j_{23} & j_{123}\\
j_4 & j & j_{234}
\end{bmatrix}
\begin{bmatrix}
j_2 & j_3 & j_{23}\\
j_4 & j_{234} & j_{34}
\end{bmatrix}
\begin{bmatrix}
j_1 & j_2 & j_{12}\\
j_{34} & j & j_{234}
\end{bmatrix}
=
\begin{bmatrix}
j_{12} & j_3 & j_{123}\\
j_4 & j & j_{34}
\end{bmatrix}
\begin{bmatrix}
j_1 & j_2 & j_{12}\\
j_3 & j_{123} & j_{23}
\end{bmatrix}.
\end{gather}
\end{subequations}
One can equivalently obtain all relations from the first one by repeadetly applying the Racah coefficients orthogonality relations
\begin{subequations}
\begin{gather}
\sum_{j_{12}}
\begin{bmatrix}
j_1 & j_2 & j_{12} \\
j_3 & j & j_{23}
\end{bmatrix}
\begin{bmatrix}
j_1 & j_2 & j_{12} \\
j_3 & j & j_{23}'
\end{bmatrix}
= \delta(j_{23},j_{23}')
\mathcal{D}(j_1,j_2|j_{12})\mathcal{D}(j_2,j_3|j_{23}),
\\[0.7em]
\sum_{j_{23}}
\begin{bmatrix}
j_1 & j_2 & j_{12} \\
j_3 & j & j_{23}
\end{bmatrix}
\begin{bmatrix}
j_1 & j_2 & j_{12}' \\
j_3 & j & j_{23}
\end{bmatrix}
= \delta(j_{12},j_{12}')
\mathcal{D}(j_1,j_2|j_{12})\mathcal{D}(j_2,j_3|j_{23}),
\end{gather}
\end{subequations}
where
\begin{equation}
\mathcal{D}(j_1,j_2|j_{12}):=
\begin{cases}
1\quad&\mbox{if } j_{12}\in \mathcal{D}(j_1,j_2)\\
0 &\mbox{if } j_{12}\not\in \mathcal{D}(j_1,j_2).
\end{cases}
\end{equation}

\subsection{Recovering the Lorentzian Ponzano--Regge model from the Hamiltonian constraint}\label{recover}

We have now all the tools to discuss the quantum Hamiltonian constraint and the Lorentzian Ponzano--Regge model \cite{Freidel:2000uq, Davids:2000kz}. The classical Hamiltonians given in \eqref{eq:general_hamiltonian_constraint} can be quantized using the quantum observables $E$, $F$ and $\widetilde F$. As discussed in section \ref{sec:holo}, we will quantize $\braket{\tau_b|\tau_b}$ as $E_b + \one$; moreover, we will choose the ordering exactly as it appears in the classical equations. The quantum Hamiltonians are given by
\begin{equation}
\begin{aligned}
\hat{H}_{abc}^{[\rangle}=& \widetilde{F}_{ac}F_{ac} +i\left( E_{ab}\widetilde{E}_{bc} - F_{ab}F_{bc} \right)\frac{F_{ac}}{E_b+\one},\quad
&
\hat{H}_{abc}^{\langle\rangle}=& \widetilde{E}_{ac}E_{ac} + i\left( E_{ab}\widetilde{F}_{bc} -F_{ab}E_{bc}\right)\frac{E_{ac}}{{E_b+\one}},\\
\hat{H}_{abc}^{\langle]}=& F_{ac}\widetilde{F}_{ac} - i\left( \widetilde{E}_{ab}E_{bc} - \widetilde{F}_{ab}\widetilde{F}_{bc} \right)\frac{\widetilde{F}_{ac}}{E_b+\one},
&
\hat{H}_{abc}^{[]}=& E_{ac}\widetilde{E}_{ac} - i\left( \widetilde{E}_{ab}F_{bc} -\widetilde{F}_{ab}\widetilde{E}_{bc}\right)\frac{\widetilde{E}_{ac}}{E_b+\one},
\end{aligned}
\end{equation}
with $\braket{abc}=\braket{234},\braket{342},\braket{423}$. Note that there is no ordering ambiguity in the fractional term, as
\begin{equation}
[F_{ac},E_b]=[\widetilde{F}_{ac},E_b]=[E_{ac},E_b]=[\widetilde{E}_{ac},E_b]=0
\end{equation}
when $a\neq b\neq c$.

We would like to show that the Lorentzian Ponzano--Regge amplitude, given by the Lorentzian Racah coefficient, is actually a solution of this constraint. To prove this we restrict ourselves to a triangular subgraph, given by the spin network
\begin{equation}
\psi(j_2,j_3,j_4):=\spinnet{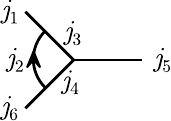};
\end{equation}
we made explicit only the dependence on $j_2$, $j_3$ and $j_4$ as these are the only legs that can be changed by $\widehat H_{abc}$.

We will consider the particular quantum Hamiltonian constraint given by 
\begin{equation}
\hat{H}^{\la]}_{342}=F_{32}\widetilde{F}_{32} - i\left( \widetilde{E}_{34}E_{42} -\widetilde{F}_{34}\widetilde{F}_{42} \right)\frac{\widetilde{F}_{32}}{E_4 +\one};
\end{equation}
all the other cases can be treated in the same way.
We want to show that $\psi$ it is annihilated by the operator $\hat{H}^{\la]}_{342}$.
The proof essentially consists in showing that the action of $\hat{H}^{\la]}_{342}$ on $\psi(j_2,j_3,j_4)$ provides a recursion relation for the Racah coefficient which is essentially the Biedenharn--Elliott relation. This was already discussed at length, for the  (undeformed and deformed) vector case \cite{Bonzom:2011hm, Bonzom:2014bua} and the spinor case \cite{Bonzom:2011nv} in the Euclidean case. We show here that this is also happening in the Lorentzian case. The main difficulty is that one needs to deal with both unitary and non-unitary representations. However as we have discussed, all these difficulties have been addressed in the previous sections. We can use our different results to prove that the spin network $\psi(j_2,j_3,j_4)$ is in the  kernel of $\hat{H}^{\la]}_{342}$.

Using the symmetries of the Racah coefficients and the actions of the $E$, $F$ and $\widetilde F$ operators from section \ref{sec:int_obs}, one finds
\begin{equation}
F_{32}\widetilde{F}_{32}\,\psi{(j_2,j_3,j_4)}=D(j_2)^2 D(j_3)^2
\racah{j_1}{j_2}{j_3}{\frac{1}{2}}{j_3-\frac{1}{2}}{j_2-\frac{1}{2}}^2
\psi(j_2,j_3,j_4).
\end{equation}
Analogously, for the other 2 parts of $H_{342}$ we have (note that each operator is acting on a different node)
\begin{equation}
\begin{split}
\widetilde{E}_{34}E_{42}\frac{\widetilde{F}_{32}}{E_4 +\one}\,\psi{(j_2,j_3,j_4)}=&-i D(j_2)^2 D(j_3)^2 \frac{D(j_4+\tfrac{1}{2})^2}{D(j_4)^2}
\racah{j_1}{j_2}{j_3}{\frac{1}{2}}{j_3-\frac{1}{2}}{j_2-\frac{1}{2}}
\racah{j_2}{\frac{1}{2}}{j_2-\frac{1}{2}}{j_4+\frac{1}{2}}{j_6}{j_4}
\\
&\times
\racah{j_3}{\frac{1}{2}}{j_3-\frac{1}{2}}{j_4+\frac{1}{2}}{j_5}{j_4}
\psi(j_2-\tfrac{1}{2},j_3+\tfrac{1}{2},j_4-\tfrac{1}{2})
\end{split}
\end{equation}
and
\begin{equation}
\begin{split}
\widetilde{F}_{34}\widetilde{F}_{42}{\widetilde{F}_{32}}{E_4 +\one}\,\psi{(j_2,j_3,j_4)}=&i D(j_2)^2 D(j_3)^2 \frac{D(j_4-\tfrac{1}{2})^2}{D(j_4)^2}
\racah{j_1}{j_2}{j_3}{\frac{1}{2}}{j_3-\frac{1}{2}}{j_2-\frac{1}{2}}
\racah{j_2}{\frac{1}{2}}{j_2-\frac{1}{2}}{j_4-\frac{1}{2}}{j_6}{j_4}
\\
&\times
\racah{j_3}{\frac{1}{2}}{j_3-\frac{1}{2}}{j_4-\frac{1}{2}}{j_5}{j_4}
\psi(j_2-\tfrac{1}{2},j_3-\tfrac{1}{2},j_4-\tfrac{1}{2}).
\end{split}
\end{equation}
By using the definitions of the Racah coefficients \eqref{eq:racah_coeff1} and the fact that, as a consequence of \eqref{eq:CG_orth2}, when $j,j'\in\mathcal{D}(j_1,j_2)$
\begin{equation}
\spinnet{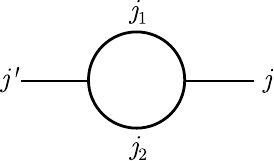}=\delta(j,j')\spinnet{id/j-j}\equiv \delta(j,j')\spinnet{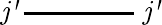},
\end{equation}
we see that
\begin{equation}
\psi(j_2,j_3,j_4)=\spinnet{hamiltonian/tet}=\sum_j\racah{j_1}{j_2}{j_3}{j_4}{j_5}{j}\spinnet{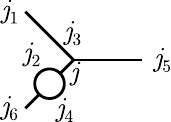}=\racah{j_1}{j_2}{j_3}{j_4}{j_5}{j_6}\spinnet{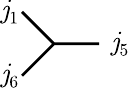}.
\end{equation}

Moreover one can prove, using the one of the Biedenharn--Elliott relations and the symmetries of Racah coefficients, that
\begin{equation}
\racah{j_1}{j_2}{j_3}{\frac{1}{2}}{J_3}{J_2}
\racah{j_1}{j_2}{j_3}{j_4}{j_5}{j_6}=\sum_{J_4}\frac{D(J_4)^2}{D(j_4)^2}
\racah{j_2}{\frac{1}{2}}{J_2}{J_4}{j_6}{j_4}
\racah{j_1}{J_2}{J_3}{J_4}{j_5}{j_6}
\racah{j_3}{\frac{1}{2}}{J_3}{J_4}{j_5}{j_4}.
\end{equation}
Substituting these results in the action of the Hamiltonian, it follows that
\begin{equation}
\hat{H}^{\la]}_{342}\,\psi(j_2,j_3,j_4)=0.
\end{equation}

\section{Relationship with \texorpdfstring{$\SU(2)$}{SU(2)} theory}

The framework we have constructed automatically describes the Euclidean case as well. Mathematically, this is a consequence of the fact that $\SU(2)$ and $\SU(1,1)$ are two real forms of the complex Lie group $\SL(2,\mathbb{C})$, i.e.
\begin{equation}
\SU(2)_\mathbb{C}\cong \SU(1,1)_\mathbb{C} \cong \SL(2,\mathbb{C}).
\end{equation}
As a consequence, the complex representations of the two groups coincide; in particular, $\SU(1,1)$ representations theory contains as a subcase all the finite-dimensional representations of $\SU(2)$ used in Euclidean LQG.
In our description, every notion at the representation theory level (spinor operators, Racah coefficients, etc.) has \emph{by design} not been restricted to unitary representations, instead allowing for any irreducible one. The only exception is the definition of the Hilbert space structure, which however, as we noted, is still valid if we restrict to finite-dimensional representations alone. Consequently, everything at the quantum level can be used to described the Euclidean case as well, by using intertwiners between finite-dimensional representations.

The same is true at the classical level. In fact, it's easy to see that for the finite-dimensional representations the spinor operators satisfy $T_\pm=\mp \widetilde{T}^\dagger_\mp$. By using the equivalent reality condition $\ov\bft_\pm = \mp \bftt_\mp$ for classical spinors the group elements \eqref{holo} becomes
\begin{equation}
g = \frac{-i}{\sqrt{\la \tau|\tau\ra}\sqrt{\la w|w\ra}}
\begin{pmatrix}
\bftp_- \bft_+ + \ov\bftp_+ \ov\bft_- & \ov\bftp_+ \ov\bft_+ - \bftp_- \bft_-\\
\bftp_+ \bft_+ - \ov\bftp_- \ov\bft_- & -\ov\bftp_- \ov\bft_+ + \bftp_+ \bft_-
\end{pmatrix}.
\end{equation}
Since
\begin{equation}
{\la \tau|\tau\ra}=|\bft_-|^2 + |\bft_+|^2 \geq 0,
\end{equation}
we have
\begin{equation}
g_{22}=\ov{g}_{11},\quad g_{21}=-\ov{g}_{12},
\end{equation}
which makes $g$ and element of $\SU(2)$. The $\su(2)$ Poisson brackets are recovered by letting $x_\pm\rightarrow -i x_\pm$; the same transformation, at the quantum level, makes the finite-dimensional representations unitary (as $\SU(2)$ representations).

\section*{Outlook}
We have constructed the generalization of the spinor approach to the 3d Lorentzian case, using tensor operators as the technical tool. This generalization has been possible thanks to the recent results of Ref. \cite{wigner_eckart}, which generalized the Wigner--Eckart theorem to the $\SU(1,1)$ case. As explained in the previous section, our framework allows to recover the $\SU(2)$ case as well: this amounts to choosing different reality condition at the classical level, while at the quantum level we recover directly the standard spinors. The Lorentzian case has nevertheless several key differences with the Euclidean case, essentially due to the fact that the $\SU(1,1)$ representation theory is more complicated than the one for $\SU(2)$: for example the spinor observables can send an intertwiner defined in terms of unitary representations to an intertwiner defined in terms of non-unitary representations. As such, they are not properly observables\footnote{There are not self-adjoint either, but this is not the case either in the Euclidean case. However, when dealing with $\SU(2)$, one can consider linear combinations which would be self-adjoint. This is not true in general with $\SU(1,1)$.}; however, one still use them to construct proper observables, such as the ones arising from the flux operator. They are still good enough to generate a solvable Hamiltonian constraint, which is not self-adjoint (nor it is in the $\SU(2)$ case). This should not be seen as an issue (since we care only about its kernel) but more as a consequence of the parametrization of the constraint using complex variables.   This constraint has the Lorentzian Ponzano--Regge amplitude in its kernel, as we would expect. We have followed here the method described in Ref. \cite{Bonzom:2011nv}, and focus on a tetrahedral spin network. Clearly this should be extended to more general graphs. There are two other main lines of future research.

\paragraph*{Extension to the 4d case.} This 3d example we considered gave insights regarding the 4d case. In order  to use the spinor (or twistor) formalism one needs first to determine the Wigner--Eckart theorem for $\SL(2,\C)$, which amounts to finding the recoupling theory between finite-dimensional non-unitary representations and the  infinite-dimensional unitary ones. The techniques developed in Ref. \cite{wigner_eckart} should prove to be useful in this case as well.   Just like for the $\SU(1,1)$ and $\SU(2)$ common description, one can expect that the spinor/twistor formalism for $\SL(2,\C)$ will be also useful to describe the Euclidean case given by $\SO(4)$.

\paragraph*{Introduction of a non-zero cosmological constant.} 3d gravity is often considered ``too'' simple when the cosmological constant vanishes. When $\Lambda\neq0$ new interesting features appear, such as the BTZ black hole in the Lorentzian case. It would therefore be interesting to generalize our spinor formalism along the lines of Ref. \cite{Dupuis:2014fya}, to investigate if new light is shed on the interesting physics happening when $\Lambda$ is non-zero.

\section*{Acknowledgement}
The authors would like to thank A. Baratin, M. Dupuis, E. Livine and S. Speziale for interesting discussions and comments.
 F. Girelli acknowledges financial support from the Government of Canada through a NSERC Discovery grant.

\appendix
\section{Dual representation}\label{app:dual_rep}

Let $\rho:G\rightarrow \mathrm{GL}(V)$ be a representation of a Lie group $G$ on an Hilbert space $V$. We define its \emph{dual representation} $\rho^*:G\rightarrow \mathrm{GL}(V^*)$ on the (continuous) dual space $V^*\cong V$ by requiring
\begin{equation}\label{eq:dual_rep1}
\braket{\rho^*(g)\alpha,\rho(g)v}=\braket{\alpha,v},\quad \forall g\in G,\quad \forall v\in V,\quad \forall\alpha\in V^*,
\end{equation}
where $\braket{\cdot,\cdot}$ is the natural pairing between $V$ and $V^*$ given by $\braket{\alpha,v}:=\alpha(v)$. Equation \eqref{eq:dual_rep1} is equivalent to
\begin{equation}\label{eq:dual_rep2}
\rho^*(g)=\left( \rho(g^{-1})\right)^*,\quad \forall g\in G,
\end{equation}
where for a linear map $f:V_1\rightarrow V_2$ we define its \emph{dual map}\footnote{Also known as \emph{transpose} in the literature.} $f^*:V_2^*\rightarrow V_1^*$ by
\begin{equation}
f^*(\alpha):=\alpha \circ f,\quad \forall \alpha\in V_2^*.
\end{equation}
By differentiating \eqref{eq:dual_rep2} we get at the Lie algebra level
\begin{equation}
\rho^*(X)=-\left( \rho(X)\right)^*,\quad \forall X\in\mathfrak{g},
\end{equation}
where with an abuse on notation we still use $\rho$ to denote the associated Lie algebra representation.

In the specific case of $\SLTR$, all continuous and finite-dimensional representations are \emph{self-dual}, i.e. isomorphic to their dual\footnote{As representations, i.e. there is an invertible intertwiner between them.}, while the dual of a discrete positive representation is isomorphic to the discrete negative representation with the same Casimir and vice versa. A possible isomorphism which works with all representation classes is
\begin{equation}
\varphi:\bra{j,m}\in V^*_j\mapsto (-1)^{m}\ket{j,-m}\in V_{j^*},
\end{equation} 
where ${j^*}$ denotes the same representation as $j$ if it is continuous or finite-dimensional, and switches between positive and negative for discrete ones.
\section{A quick overview of the proof of the Wigner--Eckart theorem}\label{proof WzE}

We provide here a summary of the proof of the Wigner--Eckart theorem.
First notice that the recoupling theory of $F_\gamma\otimes \rho_j$ is known for $j$ of any class: the proof for the infinite-dimensional classes can be found in \cite{wigner_eckart}. Assuming the decomposition exists, we can define for any $j''\in\mathcal{D}(\gamma,j)$ the vectors
\begin{equation}\label{eq:WEpsi}
\ket{\psi_{j'',m''}}:=\sum_{\mu,m}A(\gamma,\mu;j,m|j'',m'')\,T^\gamma_\mu\ket{j,m},\quad m''\in\mathcal{M}_{j''}.
\end{equation}
Using the definition of tensor operators and the Clebsch--Gordan recursion relations, one can show that
\begin{equation}
\begin{cases}
J_0\ket{\psi_{j'',m''}}=m''\ket{\psi_{j'',m''}}\\
J_\pm\ket{\psi_{j'',m''}}=C_\pm(j'',m'')\ket{\psi_{j'',m''\pm 1}},
\end{cases}
\end{equation}
so that $\ket{\psi_{j'',m''}}\propto{\ket{j'',m''}}$. The proportionality factor
\begin{equation}
N(j'',m''):=\braket{j'',m''|\psi_{j'',m''}}
\end{equation}
does not depend on $m''$: in fact, one has
\begin{equation}
\braket{j'',m''+1|J_+|\psi_{j'',m''}}=C_+(j'',m'')N(j'',m''+1)
\end{equation}
and, at the same time,
\begin{equation}
\braket{j'',m''+1|J_+|\psi_{j'',m''}}=N(j'',m'')\braket{j'',m''+1|J_+|{j'',m''}}=C_+(j'',m'')N(j'',m'')
\end{equation}
for every $m''$. Equation \eqref{eq:WEpsi} can be inverted to obtain
\begin{equation}
T^\gamma_\mu\ket{j,m}=\sum_{j''\in\mathcal{D}(\gamma,j)}\sum_{m''\in\mathcal{M}_{j''}}N(j'')B(j'',m''|\gamma,\mu;j,m)\ket{j'',m''};
\end{equation}
However, since the range of $T^\gamma_\mu$ lies in $V_{j'}$, it must necessarily be
\begin{equation}
N(j'')=0\quad \forall j''\neq j',
\end{equation}
so that ultimately
\begin{equation}
\braket{j',m'|T^\gamma_\mu|j,m}=\braket{j'\Vert T^\gamma \Vert j} B(j',m'|\gamma,\mu;j,m),
\end{equation}
with
\begin{equation}
\braket{j'\Vert T^\gamma \Vert j}:=N(j').
\end{equation}

\bibliography{bibliography}

\end{document}